\documentclass[sigconf]{acmart}
\AtBeginDocument{%
  \providecommand\BibTeX{{%
    \normalfont B\kern-0.5em{\scshape i\kern-0.25em b}\kern-0.8em\TeX}}}

\usepackage{algorithm,algpseudocode}
\usepackage{amsmath}
\usepackage{stmaryrd}
\usepackage{mdframed}
\usepackage{multicol}
\usepackage{multirow}
\usepackage{amsthm}
\usepackage{wasysym}
\usepackage{stfloats}
\usepackage{comment}
\usepackage{threeparttable}
\usepackage{url}
\usepackage{graphicx}
\usepackage{subfigure}
\usepackage{colortbl}
\usepackage{soul}
\usepackage{fancyhdr}
\usepackage{xcolor}
\definecolor{seagreen}{rgb}{0.18, 0.55, 0.34}

\def\ie{\textit{i.e.}} 
\def\etc{\textit{etc.}}
\def\eg{\textit{e.g.}}
\def\etal{\textit{et~al. }}

\newtheorem{theorem}{Theorem}

\newcommand{\boolshare}[1] {\langle #1 \rangle}
\newcommand{\addshare}[1] {\llbracket #1 \rrbracket}
\newcommand{\multshare}[1] {\llbracket #1 \rrbracket^*}

\newcommand{\rnd}{\xleftarrow{\$}}

\newcommand{\pk}{\mathsf{pk}}
\newcommand{\sk}{\mathsf{sk}}
\newcommand{\Gen}{\mathsf{Gen}}
\newcommand{\Enc}{\mathsf{Enc}}
\newcommand{\Dec}{\mathsf{Dec}}
\newcommand{\tree}{\mathcal{T}}
\newcommand{\feature}{\mathcal{X}}
\newcommand{\M}{\mathsf{M}}
\newcommand{\ArrayTree}{\mathsf{A}_{\mathcal{T}}}

\newcommand{\CT}{\mathsf{C}}

\newcommand{\BV}{\mathsf{S}}
\newcommand{\ZZ}{\mathbb{Z}}
\newcommand{\idx}{\mathsf{idx}}
\newcommand{\rdx}{\mathsf{rdx}}
\newcommand{\prf}{{F}}
\newcommand{\Fun}{\mathcal{F}}
\newcommand{\Fsprf}{\mathcal{F}_{\textsf{sprf}}}

\newcommand{\Fpre}{\mathcal{F}_{\textsf{pre}}}
\newcommand{\Fsos}{\mathcal{F}_{\textsf{sos}}}

\newcommand{\party}{{P}}
\newcommand{\sender}{\mathcal{S}}
\newcommand{\receiver}{\mathcal{R}}

\newcommand{\lmod}{\mathsf{mod}}

\newcommand{\Vbit}{ \ell_{\mathsf{v}}}
\newcommand{\Bbit}{ \ell_{\mathsf{b}}}

\newcommand{\Hmod}{n}

\newcommand{\NumBlocks}{ B}

\newcommand{\SIM}{\textsf{Sim}}
\newcommand{\ADV}{\mathcal{A}}
\newcommand{\VIEW}{\textsf{View}}
\newcommand{\HYBRID}{\textsf{H}}

\copyrightyear{2022}
\acmYear{2022}
\setcopyright{acmlicensed}%acmcopyright。rightsretained
\acmConference[ASIA CCS '22] {Proceedings of the 2022 ACM Asia Conference on Computer and Communications Security}{May 30--June 3, 2022}{Nagasaki, Japan.}
\acmBooktitle{Proceedings of the 2022 ACM Asia Conference on Computer and Communications Security (ASIA CCS '22), May 30--June 3, 2022, Nagasaki, Japan}
\acmPrice{15.00}
\acmISBN{978-1-4503-9140-5/22/05}
\acmDOI{10.1145/3488932.3517413}
\begin{document}

\fancyhead{}

\newenvironment{packed_item}{
	\begin{itemize}%[leftmargin=0.5cm]
		\setlength{\topsep}{0pt}
		\setlength{\partopsep}{0pt}
		\setlength{\itemsep}{0pt}
		\setlength{\parskip}{0pt}
		\setlength{\parsep}{0pt}
	}{\end{itemize}}

\newenvironment{packed_enum}{
	\begin{enumerate}%[leftmargin=0.5cm]
		\setlength{\topsep}{0pt}
		\setlength{\partopsep}{0pt}
		\setlength{\itemsep}{0pt}
		\setlength{\parskip}{0pt}
		\setlength{\parsep}{0pt}
	}{\end{enumerate}}

\newenvironment{packed_des}{
	\begin{description}
		\setlength{\topsep}{0pt}
		\setlength{\partopsep}{0pt}
		\setlength{\itemsep}{0pt}
		\setlength{\parskip}{0pt}
		\setlength{\parsep}{0pt}
	}{\end{description}}
	
\title{Scalable Private Decision Tree Evaluation with Sublinear Communication}


\author{Jianli Bai}
\affiliation{%
  \institution{University of Auckland}
  \city{Auckland}
  \country{New Zealand}}
\email{jbai795@aucklanduni.ac.nz}

\author{Xiangfu Song}
\affiliation{%
  \institution{National University of Singapore}
  \country{Singapore}}
\email{songxf@comp.nus.edu.sg}

\author{Shujie Cui}
\affiliation{%
  \institution{Monash University}
  \city{Melbourne}
  \country{Australia}}
\email{shujie.cui@monash.edu}

\author{Ee-Chien Chang}
\affiliation{%
  \institution{National University of Singapore}
  \country{Singapore}}
\email{changec@comp.nus.edu.sg}

\author{Giovanni Russello}
\affiliation{%
  \institution{University of Auckland}
  \city{Auckland}
  \country{New Zealand}}
\email{g.russello@auckland.ac.nz}

\begin{abstract}
Private decision tree evaluation (PDTE) allows a decision tree holder to run a secure protocol with a feature provider. By running the protocol, the feature provider will learn a classification result. Nothing more is revealed to either party. 
In most existing PDTE protocols, the required communication grows exponentially with the tree's depth $d$, which is highly inefficient for large trees. This shortcoming motivated us to design a sublinear PDTE protocol with $O(d)$ communication complexity. 
The core of our construction is a shared oblivious selection (SOS) functionality, allowing two parties to perform a secret-shared oblivious read operation from an array. 
We provide two SOS protocols, both of which achieve sublinear communication and propose optimizations to further improve their efficiency. Our sublinear PDTE protocol is based on the proposed SOS functionality and we prove its security under a semi-honest adversary. 
We compare our protocol with the state-of-the-art, in terms of communication and computation, under various network settings. The performance evaluation shows that our protocol is practical and more scalable over large trees than existing solutions.

\end{abstract}




\maketitle

\section{Introduction}
Decision trees are popular machine learning techniques for data classification. 
Due to their effectiveness and simplicity, 
decision trees have been widely adopted in various applications, such as spam filtering~\cite{bratko2006spam}, credit risk assessment~\cite{koh2006two} and disease diagnosis~\cite{podgorelec2002decision}.
Typically, there are two parties: a tree holder holding a tree model; and a feature provider holding a feature vector that needs to be classified.
However, performing the evaluation processed in such a two-party setting can lead to privacy issues. On the one hand, if the feature vectors is sent in plaintext to the model provider it might reveal individuals' information that are privacy sensitive. This might be the case in healthcare and credit risk assessment applications. On the other hand, the model is a valuable asset for the model provider. If freely accessible, it may leak sensitive information about the training data. 

Private Decision Tree Evaluation~(PDTE) protocols~\cite{kiss2019sok,wu2016privately,tai2017privacy}  address the above privacy issues.
A PDTE protocol enables two reciprocal-distrustful parties to collaboratively perform the tree evaluation without revealing any sensitive information to each other.

There are several crucial aspects when dealing with decision trees under privacy settings. For instance, in non-private decision tree evaluations, the tree is traversed from root to leaf along one path. Ideally, PDTE should also traverse the tree along on path. In this case, the total number of comparisons is linear to the depth $d$ of the tree and sublinear to the size of the tree. However, revealing the evaluation path can leak sensitive information even when the tree and the feature vector are well protected (\eg, by encryption). 
For instance, given the evaluation path, the tree holder can learn whether two feature vectors have the same range of attributes by comparing two evaluation paths; the feature provider can learn  information about the tree structure during evaluation.
In addition, even the length of a decision path can reveal significant information.
For example, if the length is unique among all decision paths, it will immediately reveal the path being evaluated. 

To protect the evaluation path, previous PDTE protocols~\cite{bost2015machine,wu2016privately,tai2017privacy,kiss2019sok} pad the tree to be complete or near-complete and run comparisons for all internal nodes to conceal the decision path information.  
As a consequence, these protocols suffer from (super) linear computation/communication complexity and are inefficient when evaluating large trees, \eg, trees containing millions of nodes~\cite{catlett1991overprvning}.

\begin{table*}[!ht]
    \footnotesize  
    \renewcommand{\arraystretch}{1.3}
	\caption{S{\upshape ummary of Existing Two-party PDTE Protocols}.}
	
% 	In our protocol, the blod parts are complexities of our HE-based protocol. 
	\label{tab::comparison:twoparty}
	\centering 
	\begin{threeparttable}%for tablefootnote, should include #usepakage threeparttable
	\begin{tabular}{|c|c|c|c|c|c|c|c|c|}
		\hline
		\hline
		\textbf{Protocol} & \textbf{Comparison} & \textbf{Communication} & \textbf{Round} & \textbf{Leakage} &\textbf{SC}&  \textbf{One-time Setup} & \textbf{Primitives}\\
		\hline
		\hline

		\hline
		Bost \etal~\cite{bost2015machine} & $\lceil m/2 \rceil$ & $O(n + m)$ & $\geq$ 6 & $m$ & \Circle & \CIRCLE & Leveled-FHE\\
		\hline
		~ Wu \etal~\cite{wu2016privately} & $2^d$ & $O(2^{d}+ (n + m)\ell)$ & 6 & $m,d$ & \Circle &\CIRCLE & AHE,OT\\
		\hline
		~ Tai \etal~\cite{tai2017privacy} & $\lceil m/2 \rceil$ & ${O}((n + m)\ell)$ & 4 & $m$ & \Circle & \CIRCLE & AHE\\
		\hline
		~ Kiss \etal~\cite{kiss2019sok}(GGG) & $d$ & ${O}(\overline{m}\ell)$ & 2 & $\overline{m},d$ & \Circle & \Circle & GC,OT\\
		\hline
		~ Kiss \etal~\cite{kiss2019sok}(HHH) & $\lceil m/2 \rceil$ & ${O}((n + m)\ell)$ & 4 & $m$ & \Circle & \CIRCLE & AHE\\
		\hline
		~ Brickell \etal~\cite{brickell2007privacy} & $d$ & ${O}((n + m)\ell)$ & 2 & $m$ & \Circle & \Circle & AHE,GC,OT\\
		\hline
        ~ Joye \etal~\cite{joye2018private} & $d$ & $O(d(\ell+n)+2^d)$ & $2d$ & $d$ & \Circle & \CIRCLE & AHE, OT\\
        \hline
		~ Tueno \etal~\cite{tueno2019private}(OT)& $d$ & $O((m+n)\ell)$ & $4d$ & $d$ & \Circle & \CIRCLE & SS,OT\\
		\hline
		~ Tueno \etal~\cite{tueno2019private}(GC)& $d$ & $O((m+n)\ell)$ & $4d$ & $d$ & \Circle & \CIRCLE & SS,GC,OT\\
		\hline
		~ Tueno \etal~\cite{tueno2019private}(ORAM)& $d$ & $O(d^4\ell)$ & $d^2+3d$ & $d$ & \CIRCLE & \CIRCLE & SS,ORAM,GC\\
		\hline
		~ Ma \etal~\cite{ma2021let} & $d$ & $O(dn\ell)$ & $2d - 1$ & $m,d$ & \LEFTcircle & \LEFTcircle & SS,GC,OT\\
		\hline
		~ Our PRF-based & $d$ & $O(dn\ell)$ & $(3r_{\prf} + 5)d$ & $m,d$ & \CIRCLE & \CIRCLE & SS,OT,PRF \\
		\hline
		~ Our HE-based & $d$ & $O(dn)$ & $8d$ & $m,d$ & \CIRCLE & \CIRCLE & SS,OT,AHE \\
		\hline
		\hline
	\end{tabular}

	\footnotesize{ \textbf{SC} represents sublinear communication, \textbf{One-time Setup} denotes the tree holder is not required to re-send the tree to the feature provider. $m$: the number of tree nodes, $\overline{m}$: the number of tree nodes in a depth-padded tree, see~\cite{kiss2019sok}, $n$: the dimension of a feature vector, $d$: the longest depth of a tree, $\ell$: the bit size of feature values, $r_{\prf}$: the number of rounds required for securely evaluating PRF $\prf$. \CIRCLE: yes, \Circle: no, \LEFTcircle: partially support.
	% $d'$: a pre-set value satisfies $d'>d$,
	% $\kappa$: computational security parameter, $c_{\prf}$: communication required by a PRF $\prf$ secure evaluation, $N$: plaintext module in Paillier encryption.
	}
    \end{threeparttable}
\end{table*}

%\smallskip 
\noindent\textbf{Techniques}. 
This paper proposes a PDTE protocol to obliviously perform decision tree evaluation without leaking the tree model, the feature values or the evaluation path. 
More importantly, our protocol has a sublinear communication complexity without relying on generic RAM-based secure computation~\cite{tueno2019private}. 

To hide which node is being accessed during decision tree evaluation, we formalize a functionality called Shared Oblivious Selection~(SOS).  
The functionality allows two parties to obliviously read an element from an array, meanwhile hiding the location and the selected value with secret sharing. 
We design two efficient SOS protocols based on different techniques. 
Our first PRF-based SOS protocol adapts Floram~\cite{doerner2017scaling}, which is a communication-efficient Oblivious RAM~(ORAM) protocol, for read-only mode.
We propose a new preprocessing technique, moving most of its communication overhead to the offline phase. 
We also design optimized masking mechanisms to make the SOS protocol more efficient. 
Our second HE-based SOS protocol explores the additive homomorphic property of Paillier encryption~\cite{paillier1999public} to eliminate two-party PRF evaluation. 
This is done by a share conversion protocol from additive arithmetic sharing to multiplicative arithmetic sharing. 
Notably, both SOS protocols achieve sublinear offline communication and constant online communication.

We design our PDTE protocols by combining a tree encoding method, the SOS functionality and secure computation. 
By initializing SOS functionality with either PRF-based or HE-based SOS protocols, we obtain two PDTE protocols with different trade-offs. 
Our PDTE protocols enjoy sublinear communication with the best security properties of existing PDTE protocols.
We prove the security against a semi-honest adversary and analyze its complexity.
As shown in Table~\ref{tab::comparison:twoparty}:
although many existing PDTE protocols can support sublinear comparisons, only the ORAM-based PDTE protocol~\cite{tueno2019private} requires sublinear communication both in the online and offline phases.
We also observe that the two-party PDTE protocol from Ma~\etal~\cite{ma2021let} only supports one single classification under standard PDTE security definition.
It is unclear how to enhance \cite{ma2021let} to support multiple invocations without re-sending new permuted encrypted trees, which essentially incurs linear (offline) communication. 
However, in some real applications like disease diagnosis, the feature provider (patient) may frequently or periodically interact with the tree holder (health center) to monitor his/her health. 
Our protocols fully support multiple PDTE queries but only need one-time setup.
The setup still needs linear communication, but the overhead will be amortized across queries.

We implemented our PDTE protocols and performed  experiments to evaluate the communication and computation performance for different trees under different network conditions. 
We also compared the performance of our PDTE with the protocols proposed in~\cite{kiss2019sok,tueno2019private,ma2019privacy}.
The results show that our PRF-based protocol reduces communication around 62$\times$ for large trees when compared to~\cite{kiss2019sok} and 0.2$\times$ than~\cite{ma2021let}. In the WAN setting with high network latency, our HE-based protocol outperforms the state-of-the-art~\cite{kiss2019sok} by 83$\times$ in terms of online computation.
When compared with \cite{tueno2019private}, our PRF-based protocol requires approximately $40\times$ less total running time while our HE-based protocol saves around $5.5\times$ total running time.
Experiments show that our PDTE protocols are practical and scalable, especially for the evaluation of large trees.

\noindent\textbf{Contributions}. Our contributions can be summarized as below: 
\begin{itemize}

\item We propose two SOS protocols that enable two parties to collaboratively and obliviously share an element from an array using only sublinear communication.

\item We propose two sublinear-communication PDTE protocols by carefully combining a modified tree encoding method, the SOS functionality and efficient secure computation techniques. 
We also propose various optimization techniques to make our protocols even more efficient. 

\item We implemented our PDTE protocols and evaluate their performance. 
The experimental results show that our protocols are practical: in particular, our PDTE protocols are scalable when evaluating large trees. 

\end{itemize}

%\smallskip 
\noindent\textbf{Paper organization}. 
We introduce background information in Section~\ref{sec:background}, and provide an overview of our techniques in Section~\ref{sec:ovewview}. 
We construct our primitives and protocols in Section~\ref{sec:construction}, and report experiments and evaluation results in Section~\ref{section:experiment}. 
We summarize related work in Section~\ref{section:related} and conclude the paper in Section~\ref{section:conclusion}.

\section{Background}\label{sec:background}

\iffalse 
\begin{figure}[!t]
	\centering
	\renewcommand{\arraystretch}{1.3}
	\includegraphics[width=2.8in]{figure/padding.pdf}
	\caption{Padding of a decision tree}
	\label{fig:padding}
\end{figure}
\fi

In this section, we introduce background information of decision tree evaluation,  cryptographic primitives and definitions used in this paper. 
Table~\ref{table::notion} shows notations used throughout this paper.
\begin{table}%[!tp]
\small
    %\footnotesize
	%\renewcommand{\arraystretch}{1.3}
	\caption{Description of Symbols \& Notations}
	\label{table::notion}
    \centering	
		\begin{tabular}{c|l }  
			\cline{1-2}
			 {\bf Symbols} & {\bf Descriptions} \\
			\hline
			$\kappa$ & computational security parameter \\
			$\lambda$ & statistical security parameter \\
			$m$ & the number of nodes in a decision tree \\
			%$n$ & dimension of feature vector \\
			$d$ & the length of the longest path in a decision tree\\
			$d'$ & a pre-defined depth satisfying $d' \ge d$\\	 
			$\mathcal{T}$ & the decision tree \\
			$\feature = (x_1,..., x_n)$ & feature vector of length $n$\\
			$\ArrayTree$ & the encoded array for a decision tree $\tree$\\
			$t/l/r/v/c$ & the threshold/left child index/right child index/\\
			 & feature ID/label of a tree node~(some non-existing \\
			 & items will be given during tree encoding)\\
			$\ell$ & the default boolean sharing bit length, \ie, $\ell = |t| =$ \\
			 &$|l| = |r| = |v| = |c| = |x_i|$ \\
			$\Vbit$ & the bit length of array elements\\
			$\Bbit$ & the bit length of PRF outputs, \eg, 64, 128, or 256 \\
			$\NumBlocks = \lceil \frac{\Vbit}{\Bbit} \rceil$ &  the number of blocks for $\Vbit$-bit element\\
			\hline
		\end{tabular}
\end{table}

\subsection{Decision Tree Evaluation}

In a decision tree $\mathcal{T}$, each non-leaf node, also called \textit{decision node}, has a threshold $t \in \mathbb {Z}_{2^\ell}$ and each leaf node, also known as \textit{classification label}, has a label value $c \in \mathbb {Z}_{2^\ell}$.
A \emph{feature vector}, \ie, a query, is the data to be classified and is denoted as $\mathcal{X} = (x_1,..., x_n) \in \mathbb{Z}_{2^{\ell}}^n$ with $n$ feature values. 

Decision tree evaluation takes a tree and a feature vector as input and outputs a label as the classification result. 
The evaluation starts from the root, and it compares the threshold $t_1$ with $x_{v(1)}$ where $v : i \in \{1, 2 , ... , m\} \rightarrow j \in \{1, 2 , ... , n\}$ is a map that determines which feature value in $\feature$ should be compared with the threshold of $i$-th node. 
We will simply use $x_{v_i}$ and $x_{v(i)}$ interchangeably throughout the paper. 
Depending on whether the comparison results in 1~($x_{v_i}<t_i$) or 0~($x_{v_i} \geq t_i$), the evaluation goes either to the left or to the right child and continues the comparison until reaching a leaf. 
We call this path from the root to a leaf as the \emph{decision path} or \emph{evaluation path} for input $\feature$. 
The \emph{depth} for a decision tree is the length of the longest path.
Without ambiguity, we use $\tree(\feature)$ to denote the classification result when using $\tree$ over feature vector $\feature$. 

\subsection{Cryptographic Primitives}

\smallskip 
\noindent\textbf{Oblivious Transfer (OT).}
OT allows a receiver to obliviously choose one out of many values from a sender~\cite{goldreich2009foundations}.
The security of OT guarantees that the receiver only learns the chosen message, and the sender has no idea which value is chosen by the receiver.
% is a fundamental primitive in secure computation, it 
OT is generally computationally expensive since it requires public-key operations.
With OT extension protocols~\cite{ishai2003extending}, it is efficient to generate (polynomially) many OTs from a small number of OTs.

\noindent\textbf{Boolean Sharing.}
%There are two kinds of secret sharing scheme throughout this paper. The first one is boolean sharing~\cite{demmler2015aby}.
We denote boolean sharing $\boolshare{x}$ as sharing of $x \in \ZZ_{2}$. For a two-party case, $\boolshare{x}$ denotes $\party_0$ holds $\boolshare{x}_0$ and $\party_1$ holds $\boolshare{x}_1$, such that $x = \boolshare{x}_0 \oplus \boolshare{x}_1$, where $\oplus$ represents bitwise XOR. 
For boolean sharing $\boolshare{x}$ and $\boolshare{y}$, $\party_0$ and $\party_1$ can compute the following operations over shares without interaction. 
Here $\oplus$ and $\cdot$ denotes addition and multiplication over $\ZZ_2$. 
\begin{packed_item}
    \item $\boolshare{z} \leftarrow \boolshare{x} \oplus \boolshare{y}$: Given $\boolshare{x}$ and $\boolshare{y}$, to compute boolean sharing of $z = x\oplus y$, $\party_0$ just computes $\boolshare{z}_0 \leftarrow \boolshare{x}_0 \oplus \boolshare{y}_0$ and $\party_1$ computes $\boolshare{z}_1 \leftarrow \boolshare{x}_1 \oplus \boolshare{y}_1$. 
    
    \item $\boolshare{z} \leftarrow \boolshare{x} \oplus c$: Given $\boolshare{x}$ and a constant $c$, to compute boolean sharing of $z = x\oplus c$, $\party_0$ just computes $\boolshare{z}_0 \leftarrow \boolshare{x}_0 \oplus c$ and $\party_1$ computes $\boolshare{z}_1 \leftarrow \boolshare{x}_1$. 
    
    \item $\boolshare{z} \leftarrow c \cdot \boolshare{x}$: Given a constant $c \in \ZZ_2$ and a boolean sharing $\boolshare{x}$, to compute boolean sharing of $z = c  \cdot x$, $\party_0$ just computes $\boolshare{z}_0 \leftarrow c  \cdot  \boolshare{x}_0$ and $\party_1$ computes $\boolshare{z}_1 \leftarrow  c  \cdot  \boolshare{x}_1$. 
\end{packed_item}

$\party_0$ and $\party_1$ need to perform an interactive protocol to compute $\boolshare{z} \leftarrow \boolshare{x}  \cdot \boolshare{y}$.
One of efficient approaches is using a Beaver Multiplication Triple~(BMT)~\cite{beaver1995precomputing}. 
A BMT $(\boolshare{a}, \boolshare{b}, \boolshare{c})$ satisfies $a  \cdot b = c$.
Suppose $\party_0$ and $\party_1$ have pre-shared a BMT $(\boolshare{a}, \boolshare{b}, \boolshare{c})$, then they can compute $\boolshare{z} \leftarrow \boolshare{x}  \cdot \boolshare{y}$ efficiently.
Specifically, $\party_0$ and $\party_1$ first compute $\boolshare{e} \leftarrow \boolshare{x}\oplus \boolshare{a}$ and $\boolshare{f} \leftarrow \boolshare{y}\oplus \boolshare{b}$, and reveal $e$ and $f$.
In the end, they compute $\boolshare{z} \leftarrow \boolshare{c} \oplus (e \cdot \boolshare{b}) \oplus (f \cdot \boolshare{a}) \oplus (e \cdot f)$, which can be done by local computation. 
BMTs over $\ZZ_2$ can be efficient prepossessed by OT extension~\cite{kolesnikov2013improved}. 

In this paper, we will mainly use boolean sharing over $\ZZ_{2}^{\ell}$ for secure computation. 
The parties share each bit of $x$ using boolean sharing, and computation is done bit-by-bit.

\noindent\textbf{Arithmetic Sharing}. 
We denote sharing a secret $x \in \ZZ_{n}$ with arithmetic sharing as $\addshare{x}$, such that $\party_0$ holds $\addshare{x}_0 \in \ZZ_n$ and $\party_1$ holds $\addshare{x}_1 \in \ZZ_{n}$ satisfying $x = \addshare{x}_0 + \addshare{x}_1~(\lmod~n)$. 
%For simplicity, We will use $x = \addshare{x}_0 + \addshare{x}_1$ interchangeably.
Arithmetic sharing is an ideal sharing semantic for computing arithmetic operations such as addition, subtraction and multiplication.
Adding arithmetic-shared $\addshare{x}$ with $\addshare{y}$ or adding $\addshare{x}$ with a constant $c \in \ZZ_n$ can be efficiently done by local computation, 
and multiplication between two shared data can be done with the help of a BMT over $\ZZ_n$.

\noindent\textbf{Share Conversion}.
Different sharing methods have their advantages/disadvantages for different kinds of computation. 
In particular, boolean sharing is friendly to the boolean circuit, including XOR, AND, \etc, while arithmetic sharing is friendly to arithmetic computation such as addition and multiplication.  
Typical computation usually contains different types of computation, thus it is better to mix-use different types of sharing forms for better efficiency.
Share conversion techniques can be used for converting between boolean sharing and arithmetic sharing:

\begin{packed_item}
    \item \emph{Boolean to Arithmetic~(B2A) conversion} : Given $\boolshare{x}$ over $\ZZ_2^{\ell}$, B2A conversion transforms $\boolshare{x}$ to its arithmetic sharing $\addshare{x}$ over $\ZZ_{2^{\ell}}$.
    B2A can be done with $\ell$ OT~\cite{demmler2015aby} in $O(1)$ round. 
    
    \item \emph{Arithmetic to Boolean~(A2B) conversion} : Given $\addshare{x}$ over $\ZZ_{2^{\ell}}$, A2B conversion transforms $\addshare{x}$ to its boolean sharing $\boolshare{x}$ over $\ZZ_2^{\ell}$. 
    A2B can be done by computing an addition circuit over $\addshare{x}_0$ and $\addshare{x}_1$ using boolean sharing~\cite{demmler2015aby, patra2021aby2} in $O(\log \ell)$ rounds.
\end{packed_item}

For efficiency reason, we mainly use boolean sharing in this paper, and deploy B2A and A2B conversion whenever necessary. 

\noindent\textbf{Distributed Oblivious RAM}.
A similar primitives related to our paper is called Distributed Oblivious RAM~(DORAM) for oblivious data access.
A famous DORAM protocol is Floram~\cite{doerner2017scaling} proposed by Doerner and Shelat, which supports both read and write over secret-shared data.
Since we only care about read operation in our setting, we show the following two read-related protocols of Floram:

\begin{packed_item}
    \item 
    \underline{$ (\sk_{\rm s}, \sk_{\rm r}, \CT) \leftarrow  \textsf{Init}(1^{\kappa}, \M)$}: 
    the protocol initializes a masked array $\CT$ from an array $\M$ of length $m$. $\sender$ obtains $\sk_{\rm s}$ and $\receiver$ obtains $\sk_{\rm r}$, and both parties locally store $\CT$.
    In details, $\sender$ holds $\boolshare{\M[i]}_{\rm s}$, chooses a PRF key $\sk_{\rm s} \rnd \{0,1\}^{\kappa}$, and sends $\boolshare{\M[i]}_{\rm s}  \oplus \prf(\sk_{\rm s}, i)$ for $i \in [0, m)$ to $\receiver$, where $\prf$ represents a PRF function. 
    Similarly, $\receiver$ holds $\boolshare{\M[i]}_{\rm r}$, chooses $\sk_{\rm r} \rnd \{0,1\}^{\kappa}$, and sends $\boolshare{\M[i]}_{\rm r} \oplus \prf(\sk_{\rm r}, i)$ for $i \in [0, m)$ to $\sender$. 
    In the end, both parties can locally compute and store $\CT$ such that $\CT[i] = \boolshare{\M[i]}_{\rm s} \oplus \boolshare{\M[i]}_{\rm r} \oplus \prf(\sk_{\rm s}, i) \oplus \prf(\sk_{\rm r}, i) = \M[i]  \oplus \prf(\sk_{\rm s}, i) \oplus \prf(\sk_{\rm r}, i)$ for $i \in [0, m)$. 
    
    \item 
    \underline{$ (\boolshare{\M[\idx]}) \leftarrow  \textsf{Read}(\sk_{\rm s}, \sk_{\rm r}, \boolshare{\idx},  \CT)$}: the protocol takes ($\sk_{\rm s}$, $\boolshare{\idx}_{\rm s}$) from $\sender$, ($\sk_{\rm r}$, $\boolshare{\idx}_{\rm r}$) from $\receiver$, and $\CT$ locally stored by both parties. In the end, the parties boolean-share $\M[\idx]$, \ie, $\sender$ receives $\boolshare{\M[\idx]}_{\rm s}$  and  $\receiver$ receives $\boolshare{\M[\idx]}_{\rm r}$. 
    In details, the parties use function secret sharing~(FSS)~\cite{boyle2015function} to share a weight-1 bit vector $\BV$ of size $m$ that satisfies $\BV[\idx] = 1$ and $\BV[i] = 0$ for all $i \ne \idx$. 
    $\sender$ computes $c_{\rm s} \leftarrow \bigoplus_{i \in [0, m)} \boolshare{\BV[i]}_{\rm s} \cdot \CT[i]$ and $\receiver$ computes $c_{\rm r} \leftarrow \bigoplus_{i \in [0, m)} \boolshare{\BV[i]}_{\rm r} \cdot \CT[i]$. 
    It is easy to check $\CT[\idx] = c_{\rm s} \oplus c_{\rm r}$.
    The parties then perform two invocations of two-party PRF evaluation to boolean-share $\boolshare{\prf(\sk_{\rm s}, {\idx})}_{\rm s}$ and $\boolshare{\prf(\sk_{\rm r}, {\idx})}_{\rm r}$, thus the parties can share $\M[\idx]$ by taking off two masks.
\end{packed_item}

\noindent\textbf{Paillier Encryption}. 
Paillier encryption scheme is a public key scheme based on Decisional Composite Residuosity problem \cite{paillier1999public}. The scheme $\mathcal{PE} =  (\textsf{Gen}, \textsf{Enc}, \textsf{Dec})$ is defined as follows:

\begin{packed_item}
	\item $\textsf{Gen}(1^\kappa):$ Take as input a security parameter $\kappa$, generate two primes $p$, $q$ (size determined by $\kappa$). Compute $N=p\cdot q$. The public key $\pk = N$ and the private key $\sk = (N, p, q)$.
	
	\item $\textsf{Enc}_{\pk}(x,r)$: Take as input a message $x \in \ZZ_N$ and a public key $\pk$, with a uniform $r \rnd \ZZ^*_N$, the ciphertext is $c := (1+N)^x \cdot r^N \bmod N^2$. We also use $\textsf{Enc}(x)$ if we do not care $r$.  
	
	\item$\textsf{Dec}_{\sk}(c)$: Take as a ciphertext $c\in \ZZ_{N^2}$ and a private key $\sk$, the decrypted plaintext is $x := \frac{(c^{\phi(N)} \bmod N^2) -1}{N} \cdot \phi(N)^{-1} \bmod N$, where $\phi(N) = (p-1)(q-1)$. 
\end{packed_item}

Paillier encryption is an additively homomorphic encryption~(AHE) scheme. At a high level, we can express the homomorphic operations as the following:
\begin{packed_item}
%\small
\item Homomorphic addition: Given two ciphertexts $c=\textsf{Enc}_{pk}(x, r)$ and $c'=\textsf{Enc}_{pk}(x', r')$, then $c_{\rm add}=c \cdot c' = ((1+N)^x\cdot r^N )\cdot ((1+N)^{x'}\cdot {r'}^N) = (1+N)^{x + x'}\cdot (rr')^N=  \textsf{Enc}_{pk}(x+x', rr')$.
\item Homomorphic multiplication with a constant: Given a ciphertext $c=\textsf{Enc}_{pk}(x,r)$ and a constant $a$, then $c_{\rm mult}=c^a = ((1+N)^x\cdot r^N)^{a} =  (1+N)^{ax}\cdot (r^{a})^N = \textsf{Enc}_{pk}(a\cdot x, r^a)$.
\end{packed_item}

%\section{Security Definition}\label{Security} 
%We design our PPDT protocol and prove its security under semi-honest security model~\cite{goldreich2009foundations}\cite{hazay2010efficient}.
%Specifically, we use the simulation-based security definition for two-party computation~\cite{lindell2017simulate}. 

\subsection{Semi-honest Security}
We design our protocols and prove their security under semi-honest security model~\cite{goldreich2009foundations,hazay2010efficient}.
%{A semi-honest adversary will faithfully follow protocol specification but try to learn more information from protocol execution. 
A protocol $\Pi$ securely computes a function $f$ under semi-honest adversary if the adversary cannot learn more information beyond what can be computed from his input and output.
%}. 
A protocol may allow the parties to learn certain leakages after execution, and we treat such leakages as a part of the output. 
Formally, let $f(x,y) = (f_0(x,y),f_1(x,y))$ be a function with inputs $x,y$ and outputs $(f_0(x,y),f_1(x,y))$.
For a two-party protocol $\Pi$ computing function $f(x,y)$, we use $\VIEW^{\Pi}_i(1^{\lambda}, x, y) = (\omega,r^i;m^{i}_1,...,m^{i}_t)$ to denote the view of the $i$-th party~($i \in \{0, 1\}$) during protocol execution where $\omega \in \{x, y\}$ depends on $i$ {, $r^i$ represents the contents of its random values, and $m^{i}_1,...,m^{i}_t$ denotes the messages received by the $i$-th party}. 

\begin{definition}
The protocol $\Pi$ securely computes $f$ for any inputs $(x,y)$ if for any party $\party_i$~($i \in \{0, 1\}$) corrupted by a semi-honest adversary $\ADV$, there exists a \emph{probabilistic polynomial time} (PPT) simulator $\SIM_i$ that can produce a simulated view that is computationally indistinguishable from $\VIEW^{\Pi}_i(1^{\lambda}, x, y)$:
\[\{\SIM_i(1^{\lambda}, x, f_i(x,y) )\} 
\overset{c}{\equiv} 
\{ \VIEW^{\Pi}_i(1^{\lambda}, x, y) \}.\]
\end{definition}

\section{Overview of Our Approach} \label{sec:ovewview}
In this section, we show our protocol setting, security requirement and techniques overview for our PDTE protocol. 

%\smallskip
\subsection{Protocol Setting and Security Guarantee} 
Our PDTE protocol is in the two-party setting in which a tree holder $\party_0$ owns a decision tree model $\tree$ and a feature provider $\party_1$ provides a feature vector $\feature = (x_1, \cdots, x_n)$.
At the end of the protocol, $\party_1$ receives a prediction $\tree(\feature)$.
We assume both parties have sufficient storage to store $\tree$ or $\feature$ locally.
In our protocol, there is no third-party involved in the computation.  

We assume an adversary is static semi-honest.
A corrupted party will strictly follow the protocol but may attempt to learn as much information as possible.
Formally, let $\Fun_{\rm DT}(\tree, \feature)$ $\rightarrow$ $(\Fun_0(\tree,$ $\feature), \Fun_1(\tree, \feature))$ be the decision tree functionality 
where a decision tree model $\tree$ and a feature vector $\feature$ is provided by $\party_0$ and $\party_1$ respectively.
In the end, the output of $\Fun_{\rm DT}$ is that $\party_i$ obtains $\Fun_i(\tree, \feature)$ for $i \in \{0,1\}$.  
A PDTE protocol ${\Pi}$ securely computes $\Fun_{\rm DT}$ if there exist PPT simulators $\SIM_0$ and $\SIM_1$ such that for any $\tree$ and $\feature$: 
\begin{align*}
&\{\SIM_0(1^{\lambda}, \tree, \Fun_0(\tree, \feature))\} 
\overset{c}{\equiv} \{ \VIEW^{\Pi}_0(1^{\lambda}, \tree, \feature) \}, \\
& \{\SIM_1(1^{\lambda}, \feature, \Fun_1(\tree, \feature))\}
\overset{c}{\equiv} 
\{ \VIEW^{\Pi}_1(1^{\lambda}, \tree, \feature) \}.
\end{align*}

For our setting, $\Fun_0(\tree,\feature)=\{\perp, \mathcal{L}_0(\feature)\}$ and $\Fun_1(\tree, \feature) = \{\tree(\feature),$ $\mathcal{L}_1(\tree)\}$. $\mathcal{L}_0$ and $\mathcal{L}_1$ denote two stateless leakage functions where $\mathcal{L}_0(\feature) = \{n\}$ and $\mathcal{L}_1(\tree) = \{m, d\}$, \ie, $\party_0$ obtains no output but only the number of features $n$ in the query whereas the output of $\party_1$ contains classification result $\tree(\feature)$, {the number of tree nodes $m$, and the length of longest decision path $d$}.

\subsection{Design Goals}
\begin{itemize}
    
    \item We aim to achieve a high security standard for our PDTE protocol. 
    The protocol should only output the classification result to the feature provider.
    Besides that, the parties only learn minimal leakages as we defined. 
    % The feature holder only learns the number of nodes~($m$) and a padded depth of decision tree~($d$).
    % The tree holder gets nothing output, but only the dimension of feature query~($n$); this is the best security we can hope from existing PDTE protocols. 
    
    \item We aim to design PDTE protocol with sublinear communication, and with concretely practical efficiency than construction from generic ORAM-based secure computation.

    \item We aim to design our PDTE protocol in a modular manner. This allows us to optimize each component for the whole PDTE protocol, which also allows us to argue the security of our PDTE protocol easily. 
    
\end{itemize}

\begin{figure}%[!t] 
	\centering 
	\includegraphics[height=1.2in]{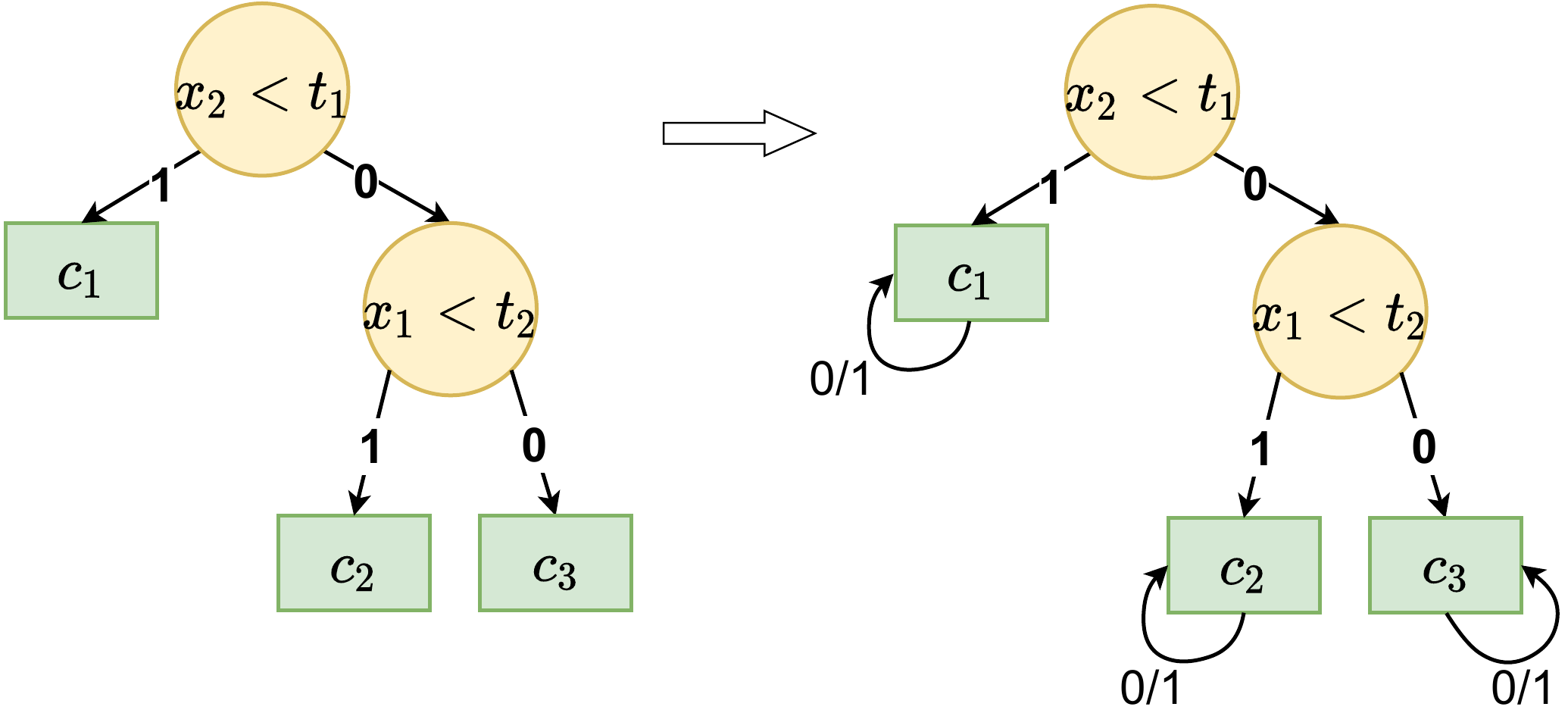} 
	\caption{{A Modified Decision Tree}} 
	\label{fig:DT-circle} 
\end{figure}

\smallskip
\subsection{Technique Overview}

\begin{figure}%[!t]
	\centering 
	\includegraphics[height=1.2in]{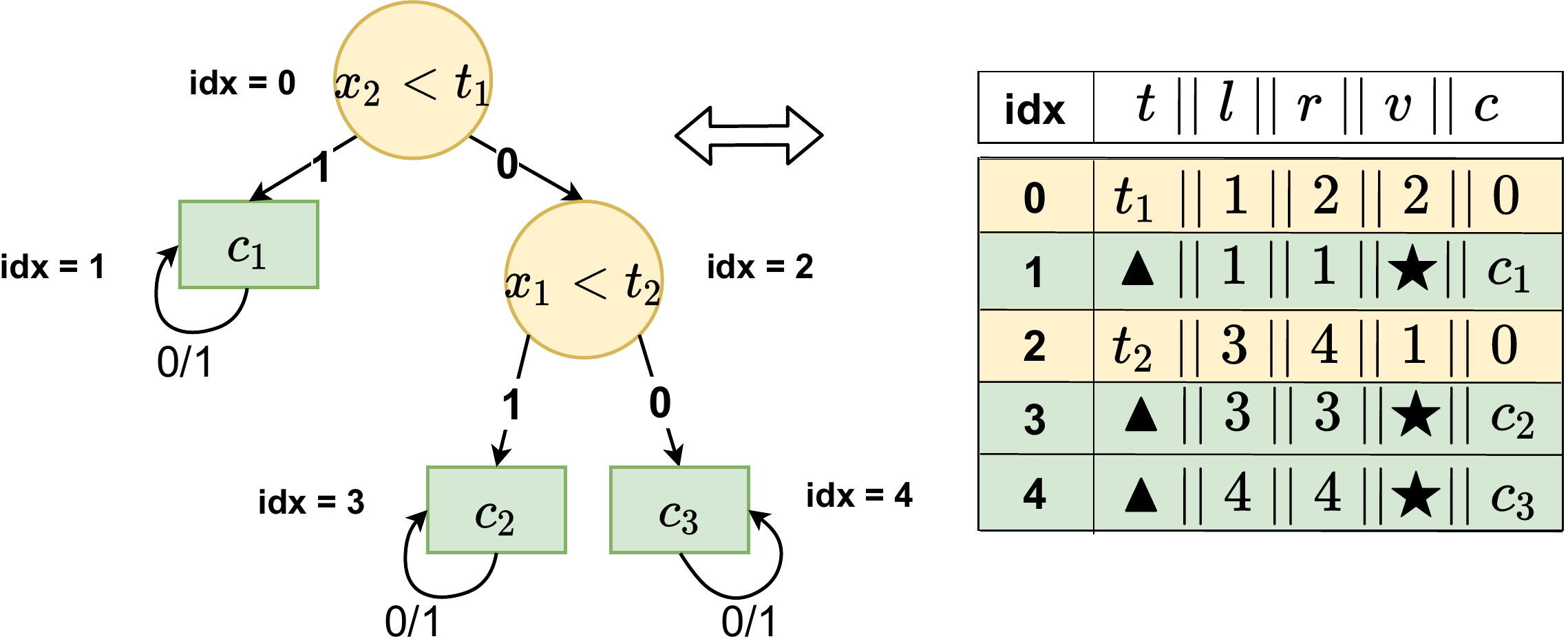} 
	\caption{Encoding A Decision Tree as an Array: `$\blacktriangle$' represents the value can set as any value, and `$\bigstar$' represents the value can be randomly selected from $[0,1,\dots,n-1]$, where $n$ is the dimension of corresponding feature vector} 
	\label{fig:tree} 
\end{figure}

\noindent\textbf{Encoding Decision Trees and Feature Vectors}. 
We follow the OAI approach~\cite{tueno2019private} to encode decision trees.
The difference is that we modify a traditional decision tree as shown in Fig.~\ref{fig:DT-circle}. 
We redirect each leaf node by setting its left and right children indexing to the leaf itself.
%As we will show later, it make our decision tree evaluation algorithm easier to hide length information.  
The modified decision tree is encoded as an array $\ArrayTree$ shown in Fig.~\ref{fig:tree}. 
A node in the tree is stored in $\ArrayTree$ in Depth First Search~(DFS) order, and we use $\ArrayTree[i]$ to store all necessary information of the $i$-th node. 
Specifically, $\ArrayTree[i]$ is constructed by five values: 1) threshold, $t$; 2) left child index, $l$; 3) right child index, $r$; 4) feature ID, $v$ and 5) classification label, $c$.
For example, the right most leaf in Fig.~\ref{fig:tree} can be represented as $\ArrayTree[4] = a||4||4||b||c_3$, where $a \rnd R$ and $b \in [0,n-1]$. 
Similarly, a feature vector from the feature provider can be naturally represented as an array $\feature$ of length $n$.

\begin{algorithm}%[!t]
\caption{$\mathsf{rst}\leftarrow \textsf{DT}(\ArrayTree, \feature)$}\label{alg:dte}
    \begin{algorithmic}[1]
    \State $\idx \leftarrow 0, \mathsf{rst} \leftarrow \perp$
    \For{$0\le i <d$} %\Comment{$d' \ge d$}
        \State \label{line3} $ t||l||r||v||c \leftarrow \ArrayTree[\idx]$
        \State \label{line4} $ b \leftarrow \feature[v] < t$
        \State \label{line5}  $ \idx \leftarrow r \oplus b\cdot (l\oplus r)$ \Comment{if $b = 1$, $\idx = l$; else $\idx = r$ }
        \State \label{line6} $ \mathsf{rst} \leftarrow c$
    \EndFor
    
    \State \Return $\mathsf{rst}$
    \end{algorithmic}
\end{algorithm}

\noindent\textbf{The Proposed Decision Tree Algorithm}. 
Given a decision tree array $\ArrayTree$ and a feature array $\feature$, we can perform decision tree evaluation over the two arrays. 
Algorithm~\ref{alg:dte} shows the algorithm, notably, it always runs $d$ iterations to output a correct classification result, independent of which path is taken.

The algorithm starts from the root node, \ie, $\idx = 0$, and it allocates a value $\mathsf{rst}$ for classification result.
In each iteration, the algorithm first selects a node $t||l||r||v||c \leftarrow  \ArrayTree[\idx]$ according to $\idx$.
From the feature ID $v$, the algorithm can select $\feature[v]$, and do a comparison $b \leftarrow \feature[v] < t$.
If $b = 1$, then set $\idx \leftarrow l$, otherwise $\idx \leftarrow r$.
In the end of each iteration, update $\mathsf{rst} \leftarrow c$. 
Due to our modification to decision tree, we can ensure that $\mathsf{rst}$ will hold a correct classification result once the evaluation reaches a leaf.
Indeed, this also ensures us to hide the length of the longest path by setting iteration number as $d'$ where $d'\ge d$. Note that setting $d'\ge d$ also hides $d$.
For simplicity, in this paper, we take $d'=d$. 

\noindent\textbf{Challenges and Solutions}. 
Things are tricky when evaluating Algorithm~\ref{alg:dte} in the secret domain. 
A basic requirement is to seal all values from both parties. 
For this part, we observe that the involved computation including comparison~(line.\ref{line4}, Algorithm~\ref{alg:dte}) and 1-out-of-2 MUX operation~(line.\ref{line5}, Algorithm~\ref{alg:dte}).
We choose boolean sharing for the underlying secret-sharing scheme because it matches well with performing bite-level secure computation.
For secure comparison and secure MUX protocol, we use existing protocols~\cite{demmler2015aby} directly. 

However, hiding intermediate values is not sufficient to get rid of all the leakages. 
Taking node selection as an example~(line.\ref{line3}, Algorithm~\ref{alg:dte}), if the secure computation leaks the memory access pattern during each iteration, the client can learn $\idx$ directly, then the client learns decision path, which is not allowed from our security requirement.
The goal here is to obliviously share $\ArrayTree[\idx]$ between parties where $\ArrayTree$ is provided by the tree holder and $\idx$ is also secret-shared, but neither party learns $\idx$ or $\ArrayTree[\idx]$\footnote{{In our protocol, ${\idx}$ is secret-shared from previous secure computation, which means neither party learns the underlying value.
Our definition of SOS functionality also ensures $\M[\idx]$ is inherently randomly shared.}
}.

\begin{figure}
	%\small
    \framebox{\begin{minipage}{0.93\linewidth}
        %\small
    		\centerline{\textbf{Functionality $\Fun^{(\M)}_{\rm sos}$}} 
    		\smallskip
    		{\bf Parameters}: Two parties denoted as $\sender$ and $\receiver$. 
    		\begin{packed_item}
    			\item \textbf{Setup}: upon receiving (\textsc{Setup}, $\M$, $\Vbit$) from $\sender$ and (\textsc{Setup}) from $\receiver$, store $\M$. 
    			
    			\item \textbf{Select}: upon receiving $(\textsc{Select}, \boolshare{\idx}_{\rm s})$ from $\sender$ and $(\textsc{Select}, \boolshare{\idx}_{\rm r})$ from $\receiver$:
    			\begin{itemize}
    			    %\item if $\mathsf{flag} = 0$, abort. 
    			    \item recover $\idx \leftarrow \boolshare{\idx}_{\rm s} \oplus \boolshare{\idx}_{\rm r}$
    			    \item $e \rnd \ZZ_{2}^{\Vbit}$, send $e$ to $\sender$ and $e \oplus \M[\idx]$ to $\receiver$
    			\end{itemize}
    		\end{packed_item}
    	\end{minipage}
	}
	\centering \caption{The Shared Oblivious Selection Functionality $\Fun_{\rm sos}^{(\M)}$
	}\label{Fun:SOS}
\end{figure}

In Fig.~\ref{Fun:SOS}, we formalize the above task as a Shared Oblivious Selection~(SOS) functionality.
A possible way to realize $\Fun_{\rm sos}^{(\M)}$ is via generic ORAM-based secure computation.
Tueno~\etal~\cite{tueno2019private} use Circuit ORAM~\cite{wang2015circuit} to design a sublinear-communication PDTE protocol. 
However, this approach requires the parties to evaluate ORAM circuit \emph{inside} secure computation, causing massive computation and communication overhead in practice.  
Our goal is to minimize the overhead by designing specialized SOS protocols. 
Our design follows from the observation that ORAM is overkill since PDTE protocols only need read operations.
Therefore, what we need is a secure computation protocol over Oblivious Read-Only-Memory~(OROM).
Such simplification allows us to design specialized SOS protocols to compute $\Fun_{\rm sos}^{(\M)}$ more efficiently. 
We also propose many optimizations to improve efficiency both asymptotically and concretely, some of them are of independent interests. 

\noindent\textbf{Put All Together}. 
We will use both SOS protocol and boolean-sharing based secure computation to evaluate our modified decision tree evaluation algorithm.
Intuitively, when performing decision tree evaluation using secure computation, all intermediate values are secret-shared between the parties, and the parties run secure computation to traverse the decision tree obliviously. 
Note that the algorithm itself does not leak length information. We further use oblivious selection to conceal access pattern leakage. 
The parties cannot learn which decision path is taken since everything is evaluated in the secret domain. 
We also propose concrete optimizations in our PDTE protocol.
We will discuss our techniques in detail in the next section.
\section{Private Decision Tree Evaluation with Sublinear Communication} \label{sec:construction}

In this section, we give our PDTE protocol with sublinear communication. 
We first propose two SOS protocols with sublinear communication under different trade-offs.
Leveraging the SOS functionality, we design our sublinear PDTE protocol. 

\begin{figure}[!hbp]
	\framebox{\begin{minipage}{0.95\linewidth}
	%\small
		{\bf Parameters}: 
		Two parties denoted as $\sender$ and $\receiver$; array length $m$; index bit length $\ell$; array element bit length $\Vbit$.
		
		\smallskip 
		{\bf [Setup]} 
		Upon receiving $(\textsc{Setup}, \M, \Vbit)$ from $\sender$ and $(\textsc{Setup}, \perp)$ from $\receiver$, $\sender$ stores $\M$ locally. 
		
	    \smallskip 
		{\bf [Select]} 
		Upon receiving $(\textsc{Select}, \boolshare{\idx}_{\rm s})$ from $\sender$ and $(\textsc{Select}, \boolshare{\idx}_{\rm r})$ from $\receiver$:   
		\begin{enumerate}
		    \item $\sender$ and $\receiver$ run a B2A conversion~\cite{demmler2015aby}, transforming $\boolshare{\idx}$ to its arithmetic form $\addshare{\idx}$ over $\ZZ_m$.%\footnote{Note that we requires $m \ge 2^\ell$ for correctness.}  
		    
		    \item $\sender$ samples $r \xleftarrow{\$} \ZZ_{2}^{\Vbit}$, computes $m$ messages $\{E_i\}_{i \in [0, m)}$ such that $E_i = r \oplus \M[i + \addshare{\idx}_{\rm s}~(\lmod~m)]$. 
		    
		    \item
		    $\sender$ and $\receiver$ invoke 1-out-of-$m$ OT functionality $\Fun_{\rm ot}$.
		    $\sender$ inputs $\{E_i\}_{i \in [0, m)}$ and $\receiver$ provides $\addshare{\idx}_{\rm r}$ as choice input. 
		    By definition of OT, $\receiver$ receives $r \oplus \M[\idx]$. 
		    
		    \item $\sender$ outputs $r$ and $\receiver$ outputs $r \oplus \M[\idx]$. %$m_{\addshare{\idx}_{\rm r}}$.
		\end{enumerate}
	\end{minipage}}
%\vspace{-0.2cm}
	\caption{Linear-communication SOS Protocol from OT \cite{tueno2019private}}
	\label{protocol:SOS_OT}	
	%\vspace{-0.4cm}
\end{figure}

\subsection{Shared Oblivious Selection Protocol}
In Fig.~\ref{protocol:SOS_OT}, we first show an SOS protocol from $1$-out-of-$m$ OT as previously done in \cite{tueno2019private}.
We will use OT-based SOS protocol over feature vectors that are usually with low dimension.
Despite being conceptually efficient and straightforward for small arrays, OT-based construction requires $O(m)$ online communication; this is prohibitively high when $m$ is large.
For oblivious selection over tree nodes, we need communication-efficient SOS protocols since, usually, a tree contains thousands to millions of nodes~\cite{catlett1991overprvning}.

\begin{figure}[!htp]
	%\small 
	\framebox{
	    \begin{minipage}{0.95\linewidth}
	    %\small
		\centerline{\textbf{Functionality $\Fun_{\rm {pre}}$}}%~\cite{keller2016mascot}}
		\smallskip
		
		{\bf Parameters}: Two parties denoted as $\sender$ and $\receiver$; weight-1 bit vector length $m$; BMT arithmetic module $\Hmod$; index bit length $\ell$. 
		\begin{packed_item}
			\item \textbf{GenWBV}: upon receiving $(\textsc{GenWBV}, \ell, m)$ from all parties: 
			\begin{itemize}
			    \item sample $\rdx \rnd \ZZ_{\ell}$, compute $\BV$ such that $\BV[\rdx] = 1$ and $\BV[j] = 0$ for all $j \ne \rdx$.
			    
			    \item sample $e \rnd \ZZ_2^{\ell}$, $ r \rnd \ZZ_2^{m}$, send $(e, r)$ to $\sender$ and $(e \oplus \rdx, \BV \oplus r)$ to $\receiver$.
			\end{itemize}
			
            \item \textbf{GenBMT}: upon receiving $(\textsc{GenBMT}, \Hmod, \addshare{a}_{\rm s}, \addshare{b}_{\rm s})$ from $\sender$ and $(\textsc{GenBMT}, \Hmod, \addshare{a}_{\rm r}, \addshare{b}_{\rm r})$ from $\receiver$: 
			\begin{itemize}
			    \item compute $c \leftarrow (\addshare{a}_{\rm s}+\addshare{a}_{\rm r})\cdot(\addshare{b}_{\rm s}+\addshare{b}_{\rm r})~(\lmod~\Hmod)$. 
			    
			    \item sample $r \rnd \ZZ_{\Hmod}$, send $-r~(\lmod~\Hmod)$ to $\sender$ and $r+c~(\lmod~\Hmod)$ to $\receiver$.
			\end{itemize}
		\end{packed_item}
	\end{minipage}
	}
	\centering \caption{The Pre-processing Functionality $\mathcal{F}_{\rm {pre}}$}
	\label{Fun:PRE}
\end{figure}
\begin{figure}[!htp]
	%\footnotesize
	\framebox{
    	\begin{minipage}{0.95\linewidth}
    	%\small
    		\centerline{\textbf{Functionality $\Fun_{\rm sprf}$}}%~\cite{keller2016mascot}}
    		\smallskip
    		{\bf Parameters}: Two parties denoted as $\sender$ and $\receiver$; PRF $F: \{0,1\}^{\kappa} \times \{0,1\}^{\ell} \rightarrow \{0, 1\}^{\Bbit}$. 
    		\begin{packed_item}
    			\item \textbf{Eval}: upon receiving (\textsc{Eval}, $\sk$, $\boolshare{\idx}_{\rm s}$) from $\sender$ and (\textsc{Eval}, $\boolshare{\idx}_{\rm r}$) from $\receiver$, compute:
     			\begin{itemize}
        			\item reconstruct $\idx \leftarrow \boolshare{\idx}_{\rm s} \oplus \boolshare{\idx}_{\rm r}$, compute $\prf(\sk, \idx)$ 
        			\item sample $r \rnd \ZZ_{2}^{\Bbit}$, send $r$ to $\sender$ and $r \oplus \prf(\sk, \idx)$ to $\receiver$
     			\end{itemize}
    		\end{packed_item}
    	\end{minipage} 
    }
	\centering \caption{The Shared PRF Functionality $\mathcal{F}_{\rm sprf}$ 
	}
	\label{Fun:SOPRF}
\end{figure}

\smallskip
\subsubsection{\textbf{Available Ideal Functionalities}}
Before describing our protocol, we introduce two ideal functionalities.
One is called preprocessing functionality $\Fun_{\textsf{pre}}$ defined in Fig.~\ref{Fun:PRE} that generates useful correlated randomnesses, including shared weight-1 bit vector~(WBV) and Beaver multiplication triple (BMT). 
Another is called two-party shared PRF functionality $\Fsprf$ in Fig.~\ref{Fun:SOPRF}. 
In particular, WBVs can be constructed from Function Secret Sharing~(FSS)~\cite{boyle2015function,boyle2016function} with sublinear communication, and BMTs can be generated from AHE or OT~\cite{demmler2015aby}. 
We summarize how to generate these correlated randomnesses in Appendix~\ref{appendix::2::c}. 
One can initialize $\Fsprf$ by evaluating a block cipher circuit using secure two-party computation. 
In the following, we will use these ideal functionalities directly; such approach, which is known as hybrid-model, is commonly used in designing secure computation protocols~\cite{goldreich2009foundations,hazay2010efficient}.

\smallskip
\subsubsection{\textbf{The PRF-based SOS Protocol}}
Our PRF-based SOS protocol in Fig.~\ref{protocol:SOS} only needs a single two-party PRF invocation regardless of the array length.
The protocol is inspired by Floram, but we propose new techniques to improve efficiency. 

\noindent\textbf{\textbf{New Pre-processing Technique}}. 
We propose a new preprocessing technique for WBVs inspired by Beaver's circuit derandomization technique~\cite{beaver1991efficient}, moving all its generation work to the offline phase. 
Let $\BV$ be a WBV over a random index $\rdx$, the parties can use $\BV$ during the online phase to share an element at location $\idx$.  
Specifically, the parties simply reveal $\delta = \rdx - \idx~(\lmod~m)$ and compute:
\begin{align*}
\boolshare{\CT[\idx]} &= \bigoplus_{i \in [0, m)} \boolshare{\BV[\delta + i ~(\lmod~m)]} \cdot \CT[i],
\end{align*}
then $\CT[\idx]$ is shared between parties as required.\footnote{$\boolshare{\BV[\delta + i~(\lmod~m)]} \cdot \CT[i]$ is computed in the secret-shared fashion, \ie, $\sender$ computes $\boolshare{\CT[\idx]}_{\rm s} = \bigoplus_{i \in [0, m)}  \boolshare{\BV[\delta + i~(\lmod~m)]}_{\rm s} \cdot \CT[i]$, and $\receiver$ computes $\boolshare{\CT[\idx]}_{\rm r} = \bigoplus_{i \in [0, m)}  \boolshare{\BV[\delta + i~(\lmod~m)]}_{\rm r} \cdot \CT[i]$.}
This derandomization technique enables the parties to pre-generate sufficient weight-1 bit vectors to trade an efficient online protocol.

Note that $\idx$ and $\rdx$ are shared in boolean form, for better efficiency, the parties first perform B2A conversion~\cite{demmler2015aby} before performing subtraction; this can be efficiently done by $\ell$ OTs in $O(1)$ round for $\ell$-bit boolean sharing~\cite{demmler2015aby}. 
Indeed, we can get rid of B2A conversion using a slightly different technique.
Specifically, when $m = 2^{k}$, the parties can simply reveal $\delta = \idx \oplus \rdx$ and compute: 
\[\boolshare{\CT[\idx]} = \bigoplus_{i \in [0, m)} \boolshare{\BV[\delta \oplus i]} \cdot \CT[i].\]
For generic cases where $2^{k-1}\le m \le 2^{k}$, $\sender$ has to pad $\M$ of size $2^k$ and randomly places elements in the padded array before encryption; this certainly needs more storage space but only doubles the storage cost at most.

\noindent\textbf{Reduce Overhead of Two-party PRF Evaluation}. 
Our protocol only needs a single mask to hide the underlying message, instead of two used in Floram~(see section~\ref{sec:background}). 
In short, Floram works for secret-shared data; it is necessary to use two masks, each for protecting a share from one party. 
For our setting, double-masking is overkill since $\sender$ already knows $\M$. We only need one mask to protect $\M$ from $\receiver$. 
As such, we change the original masking mechanism to $\CT[i] \leftarrow \M[i] \oplus \prf(\sk_{\rm s}, i)$ for $i \in [0, m)$. 
Our optimization reduces half of the two-party PRF evaluations, which can significantly improve efficiency in practice since two-party PRF evaluation contributes the main overhead to the PRF-based protocol.

In addition, we provide implementation-level optimizations. 
In particular, Floram uses AES for instantiating $\prf$. However, AES contains many AND gates, \eg, AES-128 needs 6,400 AND-gates per evaluation\footnote{\url{https://homes.esat.kuleuven.be/~nsmart/MPC/}}, which incurs significant overhead when evaluated by secure computation. 
We provide an optimized implementation from LowMC block cipher~\cite{albrecht2015ciphers}. 
LowMC is MPC-friendly, designed with much fewer AND-gates. It provides tunable options between block size, evaluation round and security level; this allows us to choose the best parameters for different scenarios.

\begin{figure}
	\framebox{\begin{minipage}{0.95\linewidth}%\small
	%\small
			{\bf Parameters}: 
			PRF $F: \{0,1\}^{\kappa} \times \{0,1\}^{\ell} \rightarrow \{0, 1\}^{\Bbit}$; index bit length $\ell$; bit length of PRF output $\Bbit$; bit length of array element $\Vbit$; number of PRF output for an element $\NumBlocks = \lceil \frac{\Vbit}{\Bbit} \rceil$; array length $m = |\M|$. 
		
		\smallskip 
		{\bf [Setup]} Upon receiving $(\textsc{Setup}, \M, \Vbit)$ from $\sender$ and $(\textsc{Setup}, \perp)$ from $\receiver$: 
		\begin{enumerate}
			\item $\sender$ samples a secret key $\sk_{\rm s} \rnd \{0,1\}^{\kappa}$ for $\prf$, and encrypts $\M$ to obtain ciphertext $\CT$ such that $\CT[i][j] = \M[i][j] \oplus \prf(\sk_{\rm s}, i||j)$ for $i \in [0, m)$ and $j \in [0, B)$.  
			
			\item $\sender$ sends $\CT$ to $\receiver$.
			$\sender$ stores $(\sk_{\rm s}, \CT)$ and $\receiver$ stores $\CT$. 
		\end{enumerate}
		
		\smallskip 
		{\bf [Select]} 
		Upon receiving $(\textsc{Select}, \boolshare{\idx}_{\rm s})$ from $\sender$ and $(\textsc{Select}, \boolshare{\idx}_{\rm r})$ from $\receiver$:   
		\begin{enumerate}
		    \item \label{SOS:select:line1} $\sender$ and $\receiver$ send $(\textsc{GenWBV}, \ell, m)$ to $\Fun_{\rm pre}$, obtain $(\boolshare{\rdx}, \boolshare{\BV})$. 
		
		    \item \label{SOS:select:line2} $\sender$ and $\receiver$ convert $\boolshare{\idx}$ and $\boolshare{\rdx}$ to arithmetic share form $\addshare{\idx}$ and $\addshare{\rdx}$ by B2A conversion~\cite{demmler2015aby}. 
		    
			\item \label{SOS:select:line3} $\sender$ and $\receiver$ compute $\addshare{\delta} = \addshare{\rdx} - \addshare{\idx}~(\mathsf{mod}~m)$, and reveal $\delta$ in clear. 
			Then they compute $\boolshare{e} = \bigoplus_{i=0}^{m-1} \boolshare{\BV[i + \delta~(\lmod~m)]} \cdot \CT[i]$. 
			
			\item $\sender$ and $\receiver$ compute $\NumBlocks$ shared indexes $\boolshare{\idx || j} = \boolshare{\idx << \lceil \log \NumBlocks \rceil} + j$ for $j \in [0, B)$, locally. 
			
			\item \label{SOS:select:line4} $\sender$ and $\receiver$ call $\Fsprf$ for each of $\NumBlocks$ blocks. For $j \in [0, \NumBlocks)$, $\sender$ inputs (\textsc{Eval}, $\sk_{\rm s}$, $\boolshare{\idx||j}_{\rm s}$) and $\receiver$ inputs (\textsc{Eval}, $\boolshare{\idx||j}_{\rm r}$) to $\Fsprf$, $\Fsprf$ sends $\boolshare{f_j}_{\rm s}$ to $\sender$ and $\boolshare{f_j}_{\rm r}$ to $\receiver$ such that  $\boolshare{f_j}_{\rm s} \oplus \boolshare{f_j}_{\rm r} = \prf(\sk_{\rm s}, \idx||j)$.
% 			For $j \in [0, \NumBlocks)$, 

			\item \label{SOS:select:line5} Let $\boolshare{f}$ be $\boolshare{f_0}||\cdots||\boolshare{f_{\NumBlocks-1}}$, $\sender$ and $\receiver$ locally compute $\boolshare{m} = \boolshare{e} \oplus \boolshare{f}$. 
		\end{enumerate}
	\end{minipage}}
%\vspace{-0.2cm}
	\caption{The PRF-based SOS Protocol}
	\label{protocol:SOS}	
	%\vspace{-0.4cm}
\end{figure}

\noindent\textbf{{Optimized Multi-block Masking Strategy}}. 
We propose an MPC-friendly masking method for $\M$ with large-size elements.
Specifically, suppose each element has a size of $\Vbit$, and the PRF output is of size $\Bbit$.
Note that when $\Bbit < \Vbit$, a single PRF output cannot mask the whole element. 
To handle the issue, we divide the element into multiple blocks and generate a mask for each.
In addition, we design a fixed-key masking strategy, which turns to be MPC-friendly.
Specifically, denote $\M[i][j]$ as $\M[i]$'s $j$-th block, $\sender$ simply encrypts the block as: 
\[
\CT[i][j] \leftarrow \M[i][j] \oplus \prf(\sk_{\rm s}, i||j),
\] 
where ${i||j} = {i} \cdot 2^{\lceil \log_2 {B} \rceil} + j$, $\NumBlocks = \lceil \frac{\Vbit}{\Bbit} \rceil$.
Since $F: \{0,1\}^{\kappa} \times \{0,1\}^{\ell} \rightarrow \{0, 1\}^{\Bbit}$ is defined over $\ell$-bits inputs, one should note that $\ell$ should be large enough, \ie, ${i} \cdot 2^{\lceil \log_2 {B} \rceil} + j < 2^{\ell}$ for all $i \in [0, m)$ and $j\in [0, \NumBlocks)$, otherwise it is possible to encounter a wrap-around issue incurring $i_1||j_1 = i_2||j_2$ and $\prf(\sk_{\rm s}, i_1||j_1) = \prf(\sk_{\rm s}, i_2||j_2)$, then $\receiver$ can easily learn:
\[
\M[i_1][j_1]\oplus \M[i_2][j_2] = \CT[i_1][j_1] \oplus \CT[i_2][j_2].
\]
As such, we require all $i||j$ are unique for $i \in [0, n)$ and $j\in [0, \NumBlocks)$. 
Indeed, setting $\lceil \log_2 {n} \rceil + \lceil \log_2 {B} \rceil \le \ell$ suffices for the goal. 
%, see Lemma~\ref{lemma:1} and its proof.
Taking a concrete example, when $\ell = 64$, $\NumBlocks = 5$, our indexing method can support oblivious selection on $2^{61}$ elements, which is already sufficient in practice.

Two benefits follow from the design. 
First, since $i$ is boolean-shared bit-by-bit and $j$ is public, sharing ${i||j}$ is essentially free: each party just cyclically left-shifts its share of ${i}$ by $\lceil \log_2 {B} \rceil$ bits, and sets the lower $\lceil \log_2 {B} \rceil$ bits to be the share of $j$, \ie, $i||j = (i << \lceil \log_2 {B} \rceil) + j$.  
Note that $j$ is publicly known to both parties, hence sharing ${j}$ is easy, \eg, $j = 0 \oplus j$.
As a result, the parties can non-interactively share ${i||j}$ for all $j \in [0, \NumBlocks)$ from the sharing of ${i}$. 
Second, since $\prf$ uses a fixed key $\sk_{\rm s}$, we can implement two-party PRF evaluation in a SIMD mode, allowing the parties to perform PRF evaluation in parallel during oblivious selection, which improves efficiency by reducing rounds. 

{
%\smallskip 
\noindent\textbf{Complexity Analysis}.
Same as Floram, the parties need to perform linear memory scan over all encrypted array elements; this is relatively cheap given highly efficient hardware nowadays. 
Besides, our protocol requires both parties to store the encrypted array locally. 
We believe the price is desirable to trade a better online communication in scenarios where multiple invocations are frequently performed between the parties. 
However, concretely, the overhead for two-party PRF evaluation can be relatively high.  
In particular, even using LowMC PRF, the parties still have to evaluate thousands of AND gates using secure computation, which can cause massive time consumption over a high-latency network; this motivates us to design a round-efficient SOS protocol. 
}

%\smallskip 
\noindent\textbf{Security}.
We have Theorem~\ref{theorem:1} to capture security of the PRF-based SOS protocol. The proof can be found in Appendices~\ref{appendix::1::a}.

\begin{theorem}\label{theorem:1}
Let $\prf$ be a secure PRF, the oblivious selection protocol in Fig.~\ref{protocol:SOS} securely computes the functionality $\Fsos^{(\M)}$ in $(\Fsprf, \Fpre)$-hybrid model under a semi-honest adversary. 
\end{theorem}

\subsubsection{\textbf{The HE-based SOS protocol}}
The prior PRF-based SOS protocol has sublinear communication, but the parties must perform a two-party PRF evaluation per oblivious selection.
Though we provide optimizations to reduce the overhead, its round complexity is relatively high given the intrinsic complexity of PRFs; this can incur considerable time consumption over a high-latency network.

\noindent\textbf{New Solution}. 
We design a sublinear SOS protocol with better round complexity by using Paillier's AHE~\cite{paillier1999public}. 
The idea is similar to the PRF-based SOS protocol, but we explore the additive homomorphic property of Paillier encryption to eliminate two-party PRF evaluation. 
Our key technique is a new share conversion protocol between additive arithmetic sharing and multiplicative arithmetic sharing over $\ZZ_{N^2}$, which is of independent interest. 

Like the PRF-based SOS protocol, $\sender$ encrypts $\M$, but uses its public key $\pk$ and sends ciphertext $\CT$ to $\receiver$.  
When selecting an element indexed by a shared index $\idx$ among $\M$, the parties still use the shared weight-1 bit vector to obliviously share $\boolshare{\CT[\idx]}$ in boolean fashion, and then convert $\boolshare{\CT[\idx]}$ to arithmetic sharing $\addshare{\CT[\idx]}$ over $\ZZ_{N^2}$ by B2A conversion.  
However, one should note that such operation can only \emph{additively} share $\CT[\idx]$: 
\[\addshare{\CT[\idx]}_{\rm s} + \addshare{\CT[\idx]}_{\rm r} = \CT[\idx]~(\lmod~N^2).\]
In order to facilitate homomorphic property of Paillier encryption, we need the ciphertext to be shared \emph{multiplicatively} over $\ZZ^*_{N^2}$: 
\[\multshare{\CT[\idx]}_{\rm s} \cdot \multshare{\CT[\idx]}_{\rm r} = {\CT[\idx]}~(\lmod~N^2).\] 
With such conversion, the parties can explore a shared decryption technique to share the encrypted message. 
Therefore, our first challenge is designing an efficient protocol to perform such conversion.

\begin{figure}[!tp]
	\framebox{
	%\small
	\begin{minipage}{0.95\linewidth}%\small
		{\bf Parameters}: 
			index bit length $\ell$; array element bit size $\Vbit \ll |N|$; array length $m =  |\M|$; computational security parameter $\kappa$; statistical security parameter $\lambda$.
		
		\smallskip 
		{\bf [Setup]} Upon receiving $(\textsc{Setup}, \M, \Vbit)$ from $\sender$ and $(\textsc{Setup}, \perp)$ from $\receiver$. Then:   
		\begin{enumerate}
			\item $\sender$ generates Paillier public/secret key pair $(\pk, \sk) \leftarrow \Gen(1^{\kappa})$.
			
			\item $\sender$ encrypts $\M$ to obtain $\CT$ such that $\CT[i] \leftarrow \Enc_{\pk}(\M[i])$ for $i \in [0, m)$.
			%Note that $|\M[i]| < \Vbit$. 
			
			\item $\sender$ sends $(\pk, \CT)$ to $\receiver$. Both $\sender$ and $\receiver$ store $\CT$ locally. 
		\end{enumerate}
		
		\smallskip 
		{\bf [Select]} 
		Upon receiving $(\textsc{Select}, \boolshare{\idx}_{\rm s})$ from $\sender$ and $(\textsc{Select}, \boolshare{\idx}_{\rm r})$ from $\receiver$. Then:   
		\begin{enumerate}
		    \item $\sender$ and $\receiver$ send $(\textsc{GenWBV}, \sigma = \lceil \log_2 m \rceil, m)$ to $\Fun_{\rm pre}$, obtain $(\boolshare{\rdx}, \boolshare{\BV})$. 
		
		    \item $\sender$ and $\receiver$ convert $\boolshare{\idx}$ and $\boolshare{\rdx}$ to arithmetic share form $\addshare{\idx}$ and $\addshare{\rdx}$ by B2A conversion~\cite{demmler2015aby}. 
		    
			\item $\sender$ and $\receiver$ compute $\addshare{\delta} = \addshare{\rdx} - \addshare{\idx}$, and reveal $\delta$ in clear. They can share  $\boolshare{x} = \boolshare{\CT[\idx]} = \bigoplus_{i=0}^{m-1} \boolshare{\BV[i + \delta]} \cdot \CT[i]$. 
			
			\item $\sender$ and $\receiver$  convert $\boolshare{x}$ to $\addshare{x}$ by B2A conversion. %, then $\sender$ and $\receiver$ convert $\addshare{x}$ to $\multshare{x}$ by $\prot_{\textsf{A2M}}$.
			
		\item The parties convert $\addshare{x}$ to its multiplicative sharing form $\multshare{x}$ over $\ZZ_{N^2}$ as follows:
		\begin{enumerate}
			\item $\sender$ samples $a \rnd \ZZ_{N^2}$ and sends $(\textsc{GenBMT}, N^2, a, 0)$ to $\Fun_{\rm pre}$.
			$\receiver$ samples $b \rnd \ZZ_{N^2}$ and sends $(\textsc{GenBMT}, N^2, 0, b)$ to $\Fun_{\rm pre}$. 
			In the end, the parties obtain sharing $\addshare{c}$ where $c = a\cdot b~(\lmod~N^2)$.
			
			\item $\sender$ samples $\gamma \rnd \ZZ_{N^2}^*$ and sends $e \leftarrow \gamma^{-1} - a~(\lmod~N^2)$ to $\receiver$. $\receiver$ sends $f \leftarrow \addshare{x}_{\rm r} - b~(\lmod~N^2)$ to $\sender$.
			
			\item $\sender$ computes $\addshare{\addshare{x}_{\rm r}\cdot \gamma^{-1}}_{\rm s} \leftarrow a \cdot f + \addshare{c}_{\rm s}~(\lmod~N^2)$ and $\receiver$ computes $\addshare{\addshare{x}_{\rm r}\cdot \gamma^{-1}}_{\rm r} \leftarrow e\cdot f + b\cdot e + \addshare{c}_{\rm r}~(\lmod~N^2)$. $\sender$ sends $\addshare{x}_{\rm s} \cdot \gamma^{-1} + \addshare{\addshare{x}_{\rm r}\cdot \gamma^{-1}}_{\rm s}~(\lmod~N^2)$ to $\receiver$.
			
			\item $\sender$ sets $\multshare{x}_{\rm s} \leftarrow \gamma~(\lmod~N^2)$. $\receiver$ sets $\multshare{x}_{\rm r} \leftarrow \addshare{x}_{\rm s}\cdot \gamma^{-1} + \addshare{x\cdot \gamma^{-1}}_{\rm s} + \addshare{x\cdot \gamma^{-1}}_{\rm r}~(\lmod~N^2)$.
		\end{enumerate}
			\item $\receiver$ samples $\beta \rnd \ZZ_{2^{\Vbit}}$, $\rho \rnd [ 0, 2^{\lambda})$ and computes $x_{\beta} \leftarrow \multshare{x}_{\rm r} \cdot \Enc_{\pk}(\beta + \rho \cdot 2^{\Vbit})~(\lmod~N^2)$.
			
			\item $\sender$ computes $\addshare{m}_{\rm s} = \Dec_{\sk}(\multshare{x}_{\rm s}\cdot x_{\beta})~(\lmod~2^{\Vbit})$, and $\receiver$ sets $\addshare{m}_{\rm r} \leftarrow -\beta~(\lmod~2^{\Vbit})$. 
			
			\item $\sender$ and $\receiver$ run A2B conversion protocol to transform $\addshare{m}$ over $\ZZ_{2^{\Vbit}}$ to its boolean sharing form $\boolshare{m}$ over $\ZZ_2^{\Vbit}$.
			
		\end{enumerate}
	\end{minipage}}
    %\vspace{-0.2cm}
	\caption{The HE-based SOS Protocol} 
	\label{protocol:SOS-he}	
	%\vspace{-0.4cm}
\end{figure}

\noindent\textbf{Additive to Multiplicative Sharing Conversion over $\ZZ_{N^2}$}. 
Given an additive sharing $\addshare{x}$ over $\ZZ_{N^2}$ where $\sender$ holding $\addshare{x}_{\rm s}$ and $\receiver$ holding $\addshare{x}_{\rm r}$, we want to convert it to its multiplicative sharing form $\multshare{x}$ satisfying $x = \multshare{x}_{\rm s} \cdot \multshare{x}_{\rm r}~(\lmod~N^2)$. %where $\multshare{x}_{\rm s}$ is in $\sender$ and $\multshare{x}_{\rm r}$ is in $\receiver$.

The idea is $\sender$ can sample a random value $\gamma \rnd \ZZ^*_{N^2}$, and $\receiver$ can recover $x\cdot \gamma^{-1}~(\lmod~N^2)$ by running a secure protocol with $\sender$.
Now the question is how to securely compute $x\cdot \gamma^{-1}~(\lmod~N^2)$. It is easy to see that: 
\[x\cdot \gamma^{-1} = \addshare{x}_{\rm s}\cdot \gamma^{-1} + \addshare{x}_{\rm r} \cdot \gamma^{-1}~(\lmod~N^2).\]
Since $\sender$ can compute $\addshare{x}_{\rm s}\cdot \gamma^{-1}~(\lmod~N^2)$ by itself, the only issue is how to share the cross term $\addshare{x}_{\rm r}\cdot \gamma^{-1}~(\lmod~N^2)$ securely.
Indeed, this can be done with the help of a BMT of special form $a\cdot b = c~(\lmod~N^2)$ where $\sender$ holds $a$, $\receiver$ holds $b$, and the parties share $c$.\footnote{Any BMT $(\addshare{a}, \addshare{b}, \addshare{c})$ can be easily transformed to a special BMT by revealing $a$ to $\sender$ and $b$ to $\receiver$, respectively.
In $\Fpre$, this can be simply done by letting $\sender$ input $(a,0)$ and $\receiver$ input $(0, b)$.} 
With the BMT, $\sender$ reveals $e \leftarrow \gamma^{-1} - a~(\lmod~N^2)$ and $\receiver$ reveals $f \leftarrow \addshare{x}_{\rm r} - b~(\lmod~N^2)$, then $\sender$ computes $\addshare{\addshare{x}_{\rm r}\cdot \gamma^{-1}}_{\rm s} \leftarrow a \cdot f + \addshare{c}_{\rm s}~(\lmod~N^2)$ and $\receiver$ computes $\addshare{\addshare{x}_{\rm r}\cdot \gamma^{-1}}_{\rm r} \leftarrow e\cdot f + b\cdot e + \addshare{c}_{\rm r}~(\lmod~N^2)$.
$\sender$ sends $\addshare{x}_{\rm s}\cdot \gamma^{-1} + \addshare{\addshare{x}_{\rm r}\cdot \gamma^{-1}}_{\rm s}~(\lmod~N^2)$ to $\receiver$. In the end,  $\sender$ holds $\gamma$ and $\receiver$ recovers $x\cdot \gamma^{-1}~(\lmod~N^2)$ by setting 
\[x\cdot \gamma^{-1} = \addshare{x}_{\rm s}\cdot \gamma^{-1} + \addshare{\addshare{x}_{\rm r}\cdot \gamma^{-1}}_{\rm s} + \addshare{\addshare{x}_{\rm r}\cdot \gamma^{-1}}_{\rm r}~(\lmod~N^2).\]
Now $x$ is multiplicatively shared between parties over $\ZZ^*_{N^2}$.

\smallskip 
\noindent\underline{\emph{Remark}}.
Note that additive sharing is over $\ZZ_{N^2}$ whereas multiplicative sharing is over $\ZZ^*_{N^2}$, we show our conversion still works and give explanation. 
%We stress that $\gamma$ is sampled by $\sender$ from $\ZZ_{N^2}^*$ rather than $\ZZ_{N^2}$. 
%In the following, we give the explanation. 
Specifically, $\ZZ_{N^2}^*$ includes all elements in $\ZZ_{N^2}$ except $\{0, p, 2p, \cdots, (p\cdot q^2-2)\cdot p, (p\cdot q^2-1)\cdot p, q, 2q, \cdots, (q\cdot p^2-2)\cdot q, (q\cdot p^2-1)\cdot q\}$. 
It is clear that if $\gamma$ is accidentally sampled from $\ZZ_{N^2} \backslash \ZZ_{N^2}^*$, there will be no way to compute $\gamma^{-1}$. 
However, the bad probability of this accident is only $ \frac{|\ZZ_{N^2} \backslash \ZZ_{N^2}^*|}{|\ZZ_{N^2}|} < \frac{p\cdot q^2 + q\cdot p^2-1}{N^2} < \frac{p\cdot q^2+ q \cdot p^2}{N^2} = \frac{1}{p} +  \frac{1}{q}$, which is negligible.
In our protocol, $\sender$ knows $p$ and $q$ so can always select $\gamma$ with inverse from $\ZZ_{N^2}^*$.
This introduces indistinguishable difference following our prior argument. 
Therefore, our conversion works over $\ZZ_{N^2}$ correctly and securely except with negligible failing/distinguishable probability. 
Besides that, we do not differentiate $\ZZ_{N^2}$ and $\ZZ^*_{N^2}$.

\noindent\textbf{Sharing Encrypted Message over $\ZZ_{2^{\Vbit}}$ Additively}.  
For a ciphertext $x = \Enc_{\pk}( m)$ that is multiplicatively shared over $\ZZ_{N^2}$, \ie, $\sender$ has $\multshare{x}_{\rm s}$ and $\receiver$ has $\multshare{x}_{\rm r}$ such that $x = \multshare{x}_{\rm s} \cdot \multshare{x}_{\rm r}~(\lmod~N^2)$, the parties can explore homomorphic property of Paillier encryption to additively share $m \in \ZZ_{2^{\Vbit}}$.
Note that $\sender$ has Paillier public/secret key pair $(\pk, \sk)$.
Specifically, $\receiver$ computes and sends a randomized ciphertext $x_{\beta} \leftarrow \multshare{x}_{\rm r} \cdot \Enc_{\pk}(\beta + \rho \cdot 2^{\Vbit})~(\lmod~N^2)$ to $\sender$, where $\beta \rnd  \ZZ_{2^{\Vbit}}$ and $\rho \rnd [ 0, 2^\lambda)$ are randomly sampled by $\receiver$.\footnote{We use Paillier's plaintext domain $\ZZ_N$ to hold messages of length $\Vbit$ where $2^{\Vbit} \ll |N|$. Here $\rho \cdot 2^{\Vbit}$ is used for statically hiding $m+\beta$, meanwhile still allows $\sender$ to compute $m+\beta~(\lmod~2^{\Vbit})$ correctly.}
$\sender$ can compute $\multshare{x}_{\rm s}\cdot x_{\beta} = \Enc_{\pk}(m+\beta+\rho \cdot 2^{\Vbit})$ and decrypt it to learn $m+\beta~(\lmod~2^{\Vbit})$. $\receiver$ has $-\beta~(\lmod~2^{\Vbit})$.
Obviously, $\sender$ and $\receiver$ finally additively share $m$ over $\ZZ_{2^{\Vbit}}$.
The parties can perform A2B conversion to transform $\addshare{m}$ over $\ZZ_{2^{\Vbit}}$ to boolean sharing $\boolshare{m}$ over $\ZZ_{2}^{\Vbit}$.

\noindent\textbf{Security}.
We have Theorem~\ref{theorem:sos-he} for the security of our HE-based SOS protocol. The proof can be found in Appendices~\ref{appendix::2::a}.

\begin{theorem}\label{theorem:sos-he}
If Paillier encryption is semantically secure and $N$ is computationally hard to factorize, the oblivious selection protocol in Fig.~\ref{protocol:SOS-he} securely computes the functionality  $\Fsos^{(\M)}$ in $(\Fun_{\rm pre})$-hybrid model under a semi-honest adversary. 
\end{theorem} 

\subsection{The Proposed PDTE Protocol}
With our data structure, decision tree algorithm and SOS protocols, it is straightforward to design our PDTE protocol modularly. 
We show the protocol in Fig.~\ref{protocol:PDTE} built on the top of an ideal SOS functionality.

\noindent\textbf{\textbf{PDTE Setup}}. 
$\party_0$ and $\party_1$ perform necessary work to setup SOS functionality for tree array $\ArrayTree$ and feature array $\feature$. 
Moreover, $\party_0$ shares $\ArrayTree[0]$~(\ie, the root node)~with $\party_1$ as the evaluation starting node. 
From the property of boolean sharing, the parties can parse the root bit-by-bit to get the sharing of all attributes $\boolshare{t}, \boolshare{l}, \boolshare{r}, \boolshare{v}$, and $\boolshare{c}$.

\noindent\textbf{\textbf{PDTE Evaluation}}. 
In each iteration, $\party_0$ and $\party_1$ first call SOS functionality $\Fun^{(\feature)}_{\rm SOS}$ to share ${\feature[v]}$.
The parties then perform a secure comparison between $\boolshare{\feature[v]}$ and $\boolshare{t}$ to compute a comparison result $\boolshare{b}$.
The evaluation can then decide which child becomes the next evaluation node by employing a MUX computation: $\boolshare{\idx} \leftarrow \boolshare{l} \oplus \boolshare{b} \cdot (\boolshare{l} \oplus \boolshare{r})$.
That is, the parties share $\boolshare{\idx} = \boolshare{l}$ if $b = 1$, otherwise $\boolshare{\idx} = \boolshare{r}$.

From the shared index $\idx$, the parities invoke $\Fun^{(\feature)}_{\rm SOS}$ to share $\ArrayTree[\idx]$. $\boolshare{t}, \boolshare{l}, \boolshare{r}, \boolshare{v}, \boolshare{c}$ are then updated correspondingly.
Besides, $\boolshare{c}$ is stored in $\boolshare{\mathsf{rst}}$ where the final classification label will stay in. 
Note that we encode a self-loop for each leaf node, thus $\boolshare{\mathsf{rst}}$ will always hold a correct classification label once the evaluation reaches a leaf node. 
Moreover, it is easy to hide length information: $\party_0$ and $\party_1$ just run $d$ iterations of evaluation. 
In the end, $\party_0$ sends $\boolshare{\mathsf{rst}}_0$ to $\party_1$, and $\party_1$ recovers ${\mathsf{rst}}$ as classification result.

The protocol runs in $O(d)$ iterations with $d$ secure comparison and MUX operations. 
If the OT-based SOS protocol is used over $\feature$ and a sublinear SOS protocol is used over $\ArrayTree$, then the total communication complexity is $O(n \ell d)$.

\begin{figure}
	\framebox{\begin{minipage}{0.95\linewidth}
	%\small
			{\bf Parameters}: Computational security parameter $\kappa$;
			$\party_0 $ provides a tree $\tree$; $\party_1$ provides a feature vector $\feature$;
		    the longest tree depth $d$.
			%$d' \ge d$ where $d$ is the maximal length of a decision path in $\tree$.  
			
		\smallskip
		{\bf [Setup]}
		\begin{enumerate}
		    \item $\party_0$ encodes its decision tree $\tree$ to an array $\ArrayTree$.  
		    
			\item $\party_0$ and $\party_1$ invoke functionality $\Fun^{(\ArrayTree)}_{\rm SOS}$. $\party_0$ sends $(\textsc{Setup}, \ArrayTree, \Vbit)$ to $\Fun^{(\ArrayTree)}_{\rm SOS}$ and $\party_1$ sends $(\textsc{Setup})$ to $\Fun^{(\ArrayTree)}_{\rm SOS}$.
			
			\item $\party_0$ and $\party_1$ invoke functionality $\Fun^{(\feature)}_{\rm SOS}$. $\party_1$ sends $(\textsc{Setup}, \feature, \ell)$ to $\Fun^{(\feature)}_{\rm SOS}$, and $\party_0$ sends $(\textsc{Setup})$ to $\Fun^{(\feature)}_{\rm SOS}$. 
			
		    \item $\party_0$ shares root node $\ArrayTree[0]$ with $\party_1$. Both parties parse $\boolshare{\ArrayTree[0]}$ as $\boolshare{t}||\boolshare{l}||\boolshare{r}||\boolshare{v}||\boolshare{c}$.
		\end{enumerate}
		
		\smallskip 
		{\bf [Evaluation]}
		\begin{enumerate}
		  \item For $i \in [1, d]$
		    \begin{enumerate}
		        \item  $\party_0$ sends $(\textsc{Eval}, \boolshare{v}_{\rm s})$ and $\party_1$  sends $(\textsc{Eval}, \boolshare{v}_{\rm r})$ to $\Fun^{(\feature)}_{\rm SOS}$. 
		        In the end, $\party_0$ and $\party_1$ share $\boolshare{\feature[v]}$.
		        
		        \item $\party_0$ and $\party_1$ run secure comparison protocol to compute  $\boolshare{b} \leftarrow \boolshare{\feature[v]} > \boolshare{t}$.
		        
		        \item $\party_0$ and $\party_1$ compute index of next tree node $\boolshare{\idx} \leftarrow \boolshare{l} \oplus \boolshare{b} \cdot (\boolshare{l} \oplus \boolshare{r})$.
		        
		        \item $\party_0$ sends $(\textsc{Eval} \boolshare{\idx}_{\rm 0})$, and $\party_1$  sends $(\textsc{Eval}, \boolshare{\idx}_{\rm 1})$ to $\Fun^{(\ArrayTree)}_{\rm SOS}$. 
		        In the end, $\party_0$ and $\party_1$ share $\boolshare{\ArrayTree[\idx]}$.
		        
		        \item parse $\boolshare{t}||\boolshare{l}||\boolshare{r}||\boolshare{v}||\boolshare{c} \leftarrow \boolshare{\ArrayTree[\idx]}$.
		        
		        \item set $\boolshare{\mathsf{rst}} \leftarrow \boolshare{c}$.
		    \end{enumerate}
		    
		    \item $\party_0$ and $\party_1$ reveal ${\mathsf{rst}}$ to $\party_1$ as output. 
		\end{enumerate}
	\end{minipage}}
%\vspace{-0.2cm}
	\caption{Our PDTE Protocol}
	\label{protocol:PDTE}
	%\vspace{-0.4cm}
\end{figure}

\noindent\textbf{Optimization 1 - Reduce SOS Invocations}. 
We can reduce the number of SOS invocations by exploring a data locality property in decision tree evaluation. 
Our observation is that tree evaluation will only go from a parent to one of its children.
Therefore, we can pack the parent with its children together as a bigger node to reduce invocations of oblivious selection. 
We call the packed node as a \emph{cluster}. 
If the parent node is a leaf, $\sender$ needs to allocate two dummy nodes to make the cluster's size indistinguishable from others.
The parties then use oblivious selection to share the desired cluster between parties.
Since our SOS protocol supports SIMD mode, the parties can share a cluster by only one invocation, whereas it requires two in the original protocol. 
In this way, we reduce PDTE evaluation invocation from $d$ to $\lceil d/2 \rceil$.
We can generalize the idea to pack a parent node with its descendants in the following $q$ layers, reducing invocations from $d$ to $\lceil d/q \rceil$.

The remaining issue is how to traverse within a cluster obliviously.
This can be done by $q$ MUX operations, and each is over two smaller sub-trees.
However, the total communication for traversing within a cluster will be $O(2^q)$. 
In practice, we can set $q = 2$ or 3 to reduce 50\% or 67\% rounds from SOS protocol while not increasing communication too much. 

\noindent\textbf{Optimization 2 - Reduce Local Computation}. 
% As we can see, our PDTE can be applied to both complete and sparse tree without any difference.
In our PDTE protocol, each oblivious selection causes a linear scan over the whole decision tree, incurring $O(d\cdot m)$ computation in total.
We can reduce the overhead when $\tree$ is a complete tree.
That is, instead of performing the scan over all nodes, the parties only need to run oblivious selection over $i$-th layer of $\tree$ for the $i$-th iteration of evaluation. 
Therefore, the total local computation from SOS will only be $O(m)$.
%For complete tree, this optimization cause no more leakage.
Note that we can not use this optimization directly over sparse trees; otherwise, $\receiver$ can learn the tree structure of each layer, \eg, number of nodes of each layer.
Nevertheless, we can always transform a non-complete tree into a complete one by padding dummy nodes and then we can optimize the padded tree. 
In practice, padding is cost-effective for those near-complete trees but not for sparse trees.

\noindent\textbf{Security}.
We have Theorem~\ref{theorem:2} towards security of our PDTE protocol. The proof can be found in Appendices~\ref{appendix::3::a}.

\begin{theorem}\label{theorem:2}
The PDTE protocol in Fig.~\ref{protocol:PDTE} securely computes the functionality $\Fun_{\rm DT}$ in $(\Fun^{(\ArrayTree)}_{\rm SOS}, \Fun^{(\feature)}_{\rm SOS})$-hybrid model against semi-honest adversary.
\end{theorem}
\section{Experiment}\label{section:experiment}
In this section, we report the concrete efficiency of our PDTE protocol. 
We implement the protocol in C++ under ABY framework~\cite{demmler2015aby}. 

\subsection{Experiment Setup}
We run our experiment on a desktop PC equipped with Intel(R) Core™ i9-9900 CPU at 3.10~GHz × 16 running Ubuntu 20.04 LTS and 32~GB of memory. % that supports Intel's AES-NI for fast AES operations. 
We use Linux \textsf{tc} tool to simulate local-area network~(LAN, RTT: 0.1~ms, 1~Gbps), metropolitan-area network~(MAN, RTT: 6~ms, 100~Mbps) and wide-area network~(WAN, RTT: 80~ms, 40~Mbps).
We set the computational security parameter $\kappa = 128$ and statistical security parameter $\lambda = 40$. 
As in prior work, we set the bit length to $\ell = 64$. For AHE, the plaintext module is $|N| = 2048$.
We implement the involved secure computation using GMW~\cite{micali1987play} protocol over boolean sharing as default. The times reported are averaged over ten trials.

\subsection{Tree Parameters}
We evaluate our protocols on 8 representative datasets from UCI repository\footnote{\url{https://archive.ics.uci.edu/ml}} as listed in Table~\ref{tab::paramters}. 
To compare with~\cite{kiss2019sok, ma2021let}, we directly use their used decision trees \textit{wine, Linnerud, breast, digits, diabetes} and \textit{Boston} which are trained using codes from~\cite{kiss2019sok}. 
We additionally train two trees, one is a deep-but-sparse tree \textit{spmabase} and another is a density tree \textit{MNIST} with a high-dimensional vector.

\begin{table}[!hp]
\footnotesize
\renewcommand{\arraystretch}{1.1}
\caption{Tree Parameters}\label{tab::paramters}
\centering
%\scalebox{0.9}{%minimize the whole table
	\begin{tabular}{|c|c|c|c|}
	\hline
	\textbf{Decision Tree} & \textbf{Feature Dimension} $n$ & \textbf{Depth} $d$ & $\#$(\textbf{Nodes}) $m$\\
	\hline
% 	\rowcolor{gray!40}
    wine & 7 & 5 & 23\\
	\hline
% 	\rowcolor{gray!40}
    Linnerud & 3 & 6 & 39 \\
	\hline
% 	\rowcolor{gray!40}
	breast & 12 & 7 & 43 \\
	\hline
% 	\rowcolor{gray!40}
    digits & 47 & 15 & 337\\
	\hline
% 	\rowcolor{gray!40}
	spambase & 57 & 17 & 171 \\
	\hline
% 	\rowcolor{gray!40}
	diabetes & 10 & 28 & 787 \\
	\hline
% 	\rowcolor{gray!40}
	Boston & 13 & 30 & 851 \\
	\hline
% 	\rowcolor{gray!40}
	MNIST & 784 & 20 & 4179 \\
	\hline
    \end{tabular}
\end{table}

\begin{figure}%[!tp]
\centering
    \subfigure[Online Communication] {\label{fig:online-communication:a} \includegraphics[height=1.2in,width=1.6in]{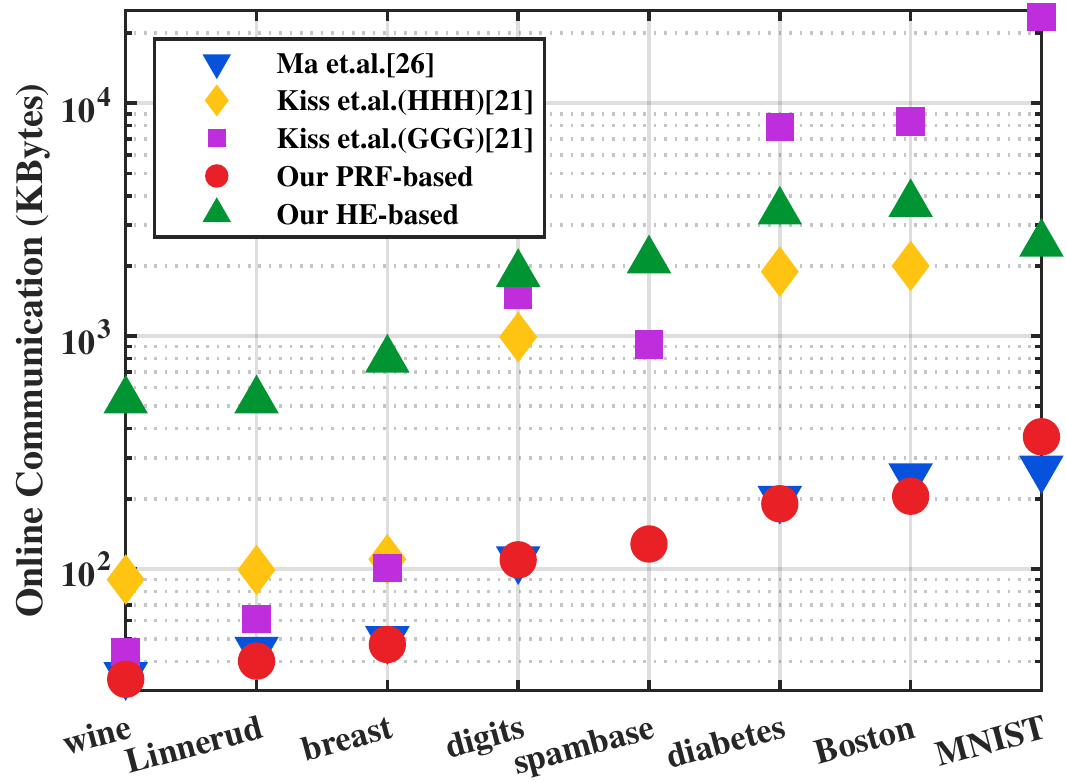}}
    \subfigure[Offline Communication] {\label{fig:online-communication:b} \includegraphics[height=1.2in,width=1.6in]{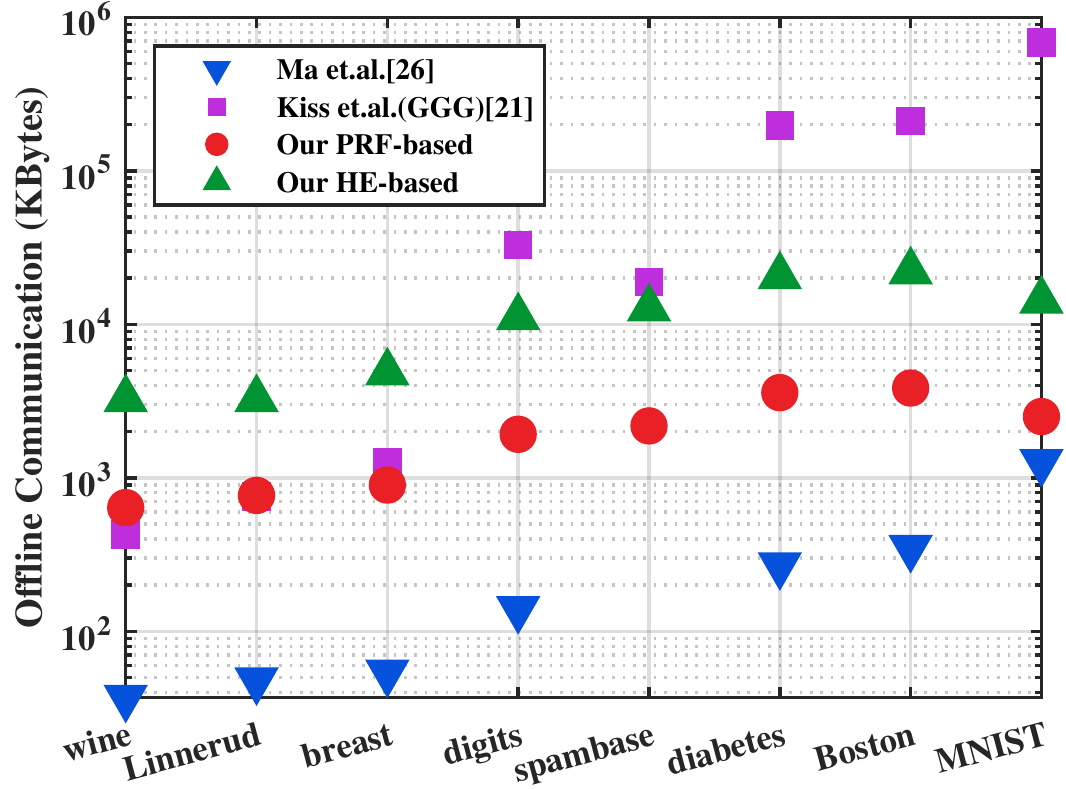}}
    % \subfigure[Total online communication cost] {\label{fig:onlineCommunication:c} \includegraphics[height=1.4in,width=1.8in]{figure/TotalCommunicationComparison.pdf}}
\caption{Online and Offline Communication Cost. Note that the $y$-axis is in logarithm scale.}
\label{fig:online-offline}
\end{figure}

\subsection{Performance Evaluation}
In this section, we report the efficiency of our protocol.
We first test PDTE protocols in communication and running time under different network settings and compare them with state-of-the-art PDTE protocols. 
Then we discuss trade-off by exploring the modular design of our PDTE design and give recommendations for different scenarios.  
Last we report performances over large synthetic deep trees to show the scalability of our PDTE protocols. 

We mainly compare our protocols with three representative PDTE works~\cite{kiss2019sok,tueno2019private,ma2021let}. Kiss~\etal~\cite{kiss2019sok} divide a PDTE protocol into three sub-protocols: feature selection, comparison and path evaluation and use either Garbled Circuit~(GC) or AHE to instantiate them. We select~\cite{kiss2019sok} because this work is the most summative in linear-cost PDTE protocols. We compare our work with their GGG and HHH since the former is computation-friendly and the latter is communication-friendly. Other two PDTE works are both sublinear-cost~\cite{tueno2019private,ma2021let} protocols, similar to ours. 
But only the ORAM-based PDTE protocol in \cite{tueno2019private} is truly sublinear-communication.

\begin{figure*}[!htp]
\centering
\subfigure[UCI datasets on LAN (1Gbps/0.1ms)] {\label{fig:onlineRuntime:a}
\includegraphics[height=1.20in,width=1.62in]{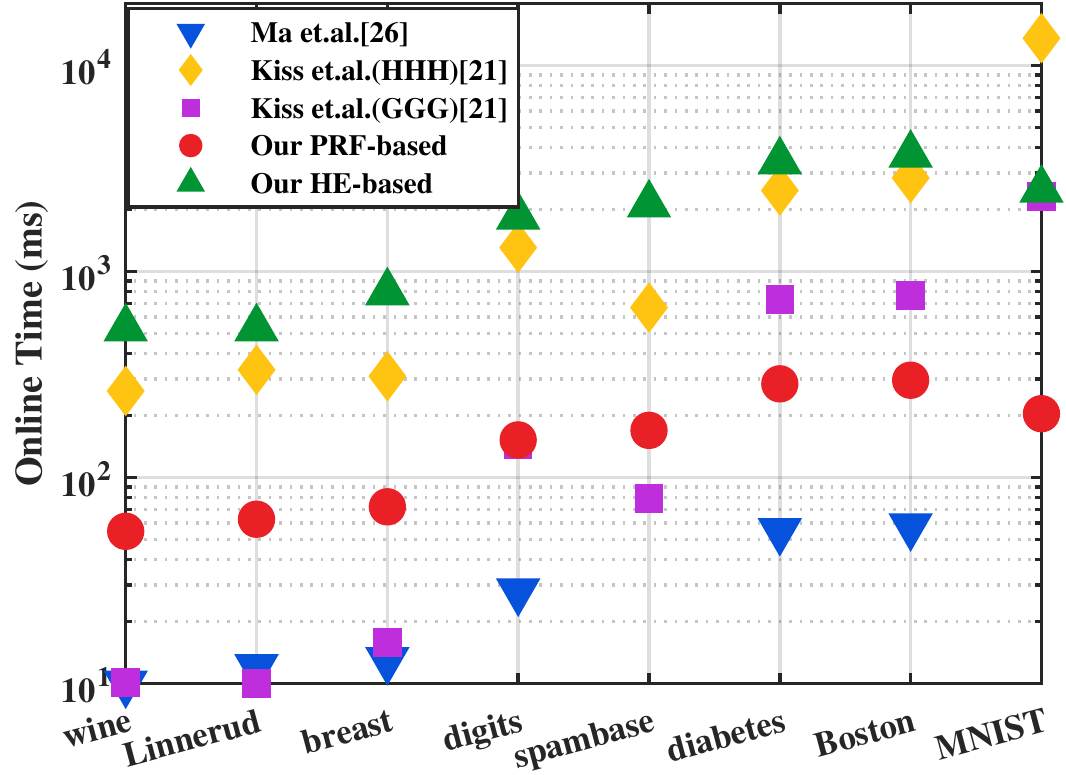}}
\subfigure[UCI datasets on MAN (100Mbps/6ms)] {\label{fig:onlineRuntime:b}
\includegraphics[height=1.22in,width=1.62in]{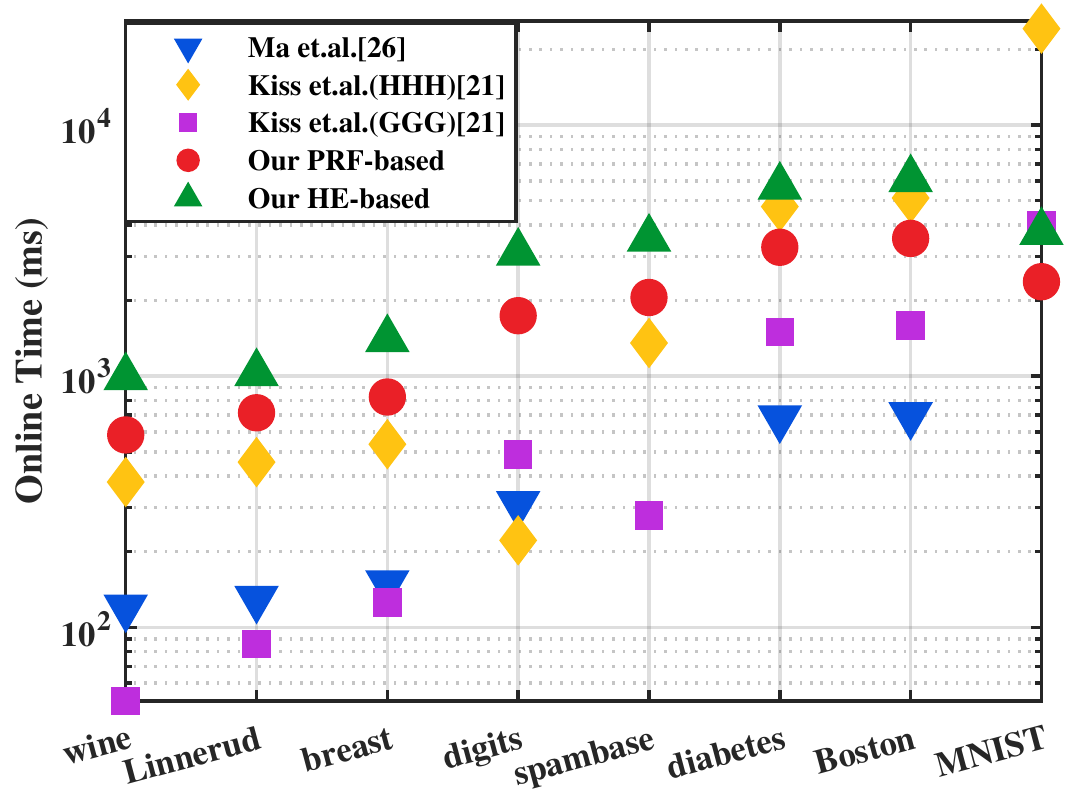}}
\subfigure[UCI datasets on WAN (40Mbps/80ms)] {\label{fig:onlineRuntime:c}
\includegraphics[height=1.224in,width=1.62in]{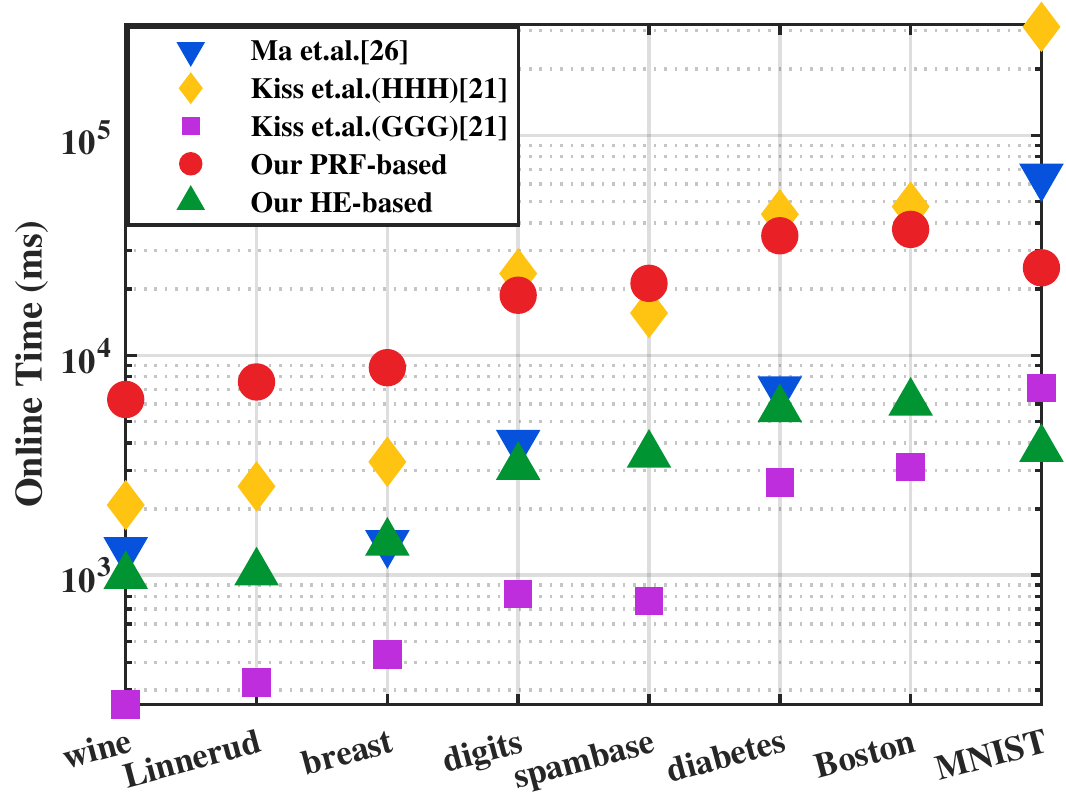}}
\caption{Online Runtime in LAN/MAN/WAN Setting. Note that the $y$-axis is in logarithm scale.}
\label{fig:onlineRuntime}
\end{figure*}

%\smallskip 
\noindent\textbf{PDTE Communication}.  
Fig.~\ref{fig:online-offline} details the communication consumption of our two PDTE protocols, and the comparison with~\cite{kiss2019sok,ma2021let}.
As we can see, our PRF-based construction~(with LowMC as the PRF instantiation) requires the least communication among these PDTE protocols and is slightly better than~\cite{ma2021let}.
Our HE-based PDTE protocol requires more communication than PRF-based one.
The main reason is that the ciphertext size evaluated in each round in our PRF-based protocol is smaller than that in our HE-based protocol. In the latter, for example, the ciphertext size is set to be $|N^2| = 4096$.
Although they all enjoy constant communication complexity per selection, the constant factor is much higher in HE-based protocol. 
GGG also shows better online communication performance than our HE-based protocol when trees are small. It is reasonable because GGG shifts all GC generation to the offline phase, resulting in efficient online efficiency independent of the tree size. However, GGG needs $n\ell$ 1-out-of-2 OT to perform oblivious feature selection. 
Thus, compared with GGG, our HE-based protocol shows less online communication cost when it comes to trees with a high-dimensional feature vector, like MNIST.

Towards offline communication, our two constructions both lie between GGG and the construction of Ma~\etal~\cite{ma2021let}. The protocol in~\cite{ma2021let} enjoys the lowest offline communication cost for small trees. However, their offline communication cost increases with the size of the tree since the tree in their protocol should be re-sent before each evaluation. Thus, as the tree size $m$ increases, our offline communication overhead will be outpaced by~\cite{ma2021let}. GGG costs the most even when the tree is medium-sized in Table~\ref{tab::paramters}, \ie, digits. As we discussed before, GGG can enjoy a better online communication cost by moving major communication to offline. However, as shown in Fig.~\ref{fig:online-communication:b}, the price is high, referring to big trees.

\noindent\textbf{PDTE Running Time}. 
We report the online running time of our PDTE protocols under different network settings~(LAN, MAN, WAN) in Fig.~\ref{fig:onlineRuntime}. The reported runtimes of protocol~\cite{ma2021let} are read from their paper. In this test, we use LowMC to instantiate the involved PRF in our PRF-based protocol. 

In the LAN setting, our HE-based protocol needs the most running time except for MNIST. It is clear to see that the running time of GGG and HHH grow with the tree size. Our HE-based protocol shows less computation when treating MNIST ($m = 4179$) than Boston ($m = 851$). This is because the depth of MNIST is smaller than Boston. Our PRF-based protocol lies between Ma \etal~\cite{ma2021let} protocol and HHH protocol. There is no doubt that protocol in~\cite{ma2021let} is the most efficient protocol to date under the LAN setting. This is because the most expensive operations in their protocol are OT and GC. Yet, in our PRF-based protocol, the most costly are secure LowMC evaluation. However, our PRF-based protocol still outruns linear protocols GGG and HHH by $24 \times$ to $65 \times$, respectively, for MNIST.

Our two protocols are slightly less efficient than HHH and require around $100 \times$ costs than GGG~\cite{ma2021let}, especially when the tree is tiny. One reason is that our work is based on GMW who is highly influenced by latency. Thus our PRF-based protocol runs an order slower when the latency increases from 0.1ms to 6ms. This ``negative" property appears on our HE-based protocol as well but in a mild influence since our HE-based protocol removes the costly LowMC operation. One should notice that our two protocols become slightly better than GGG and save at least $5 \times$ running times than HHH, for MNIST. With the increase of size/depth of the tree, our sublinear protocols will be more competitive.

We can also observe that as the network latency increases, our HE-based PDTE protocol gradually surpasses PRF-based one, \ie, $6 \times$ to $7 \times$ faster in the WAN setting.
The reason is that two-party PRF evaluation in the PRF-based protocol involves higher rounds than the HE-based construction, which causes significant time consumption over the high-latency network.

\begin{figure*}%[!hp]
\centering
\subfigure[Trade-off under LAN (1Gbps/0.1ms)] {\label{fig:tradeoff:a}
\includegraphics[height=1.233in,width=1.63in]{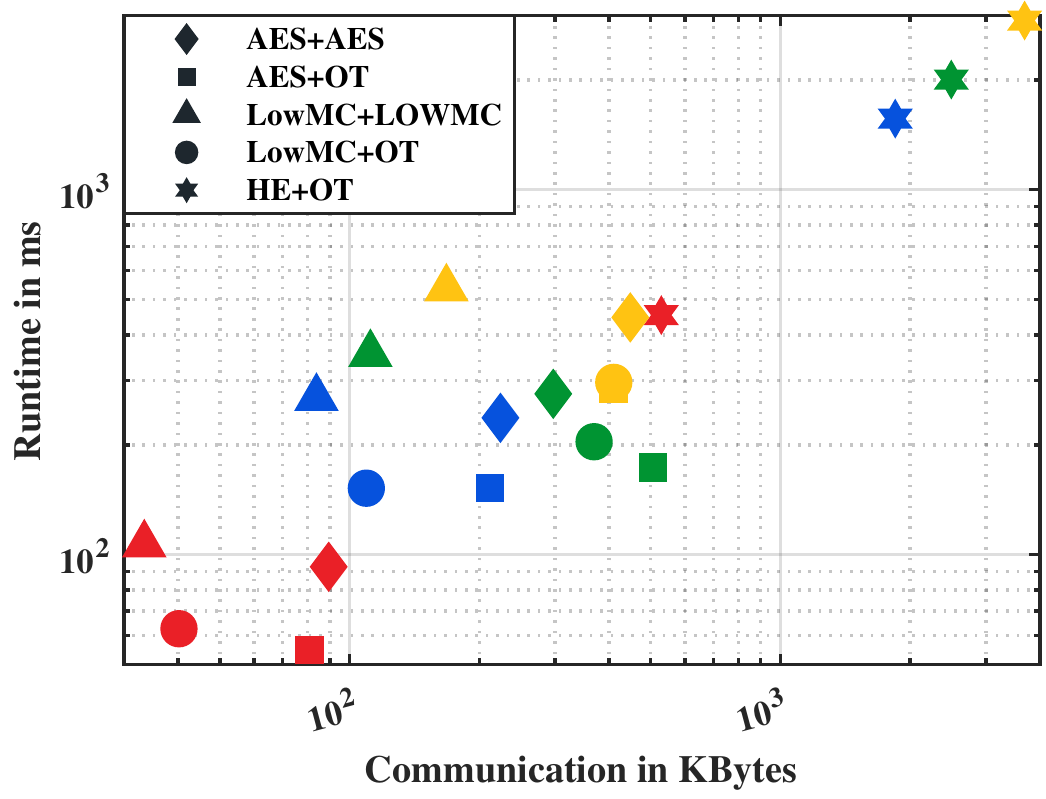}}
\subfigure[Trade-off under MAN (100Mbps/6ms)] {\label{fig:tradeoff:b}
\includegraphics[height=1.23in,width=1.63in]{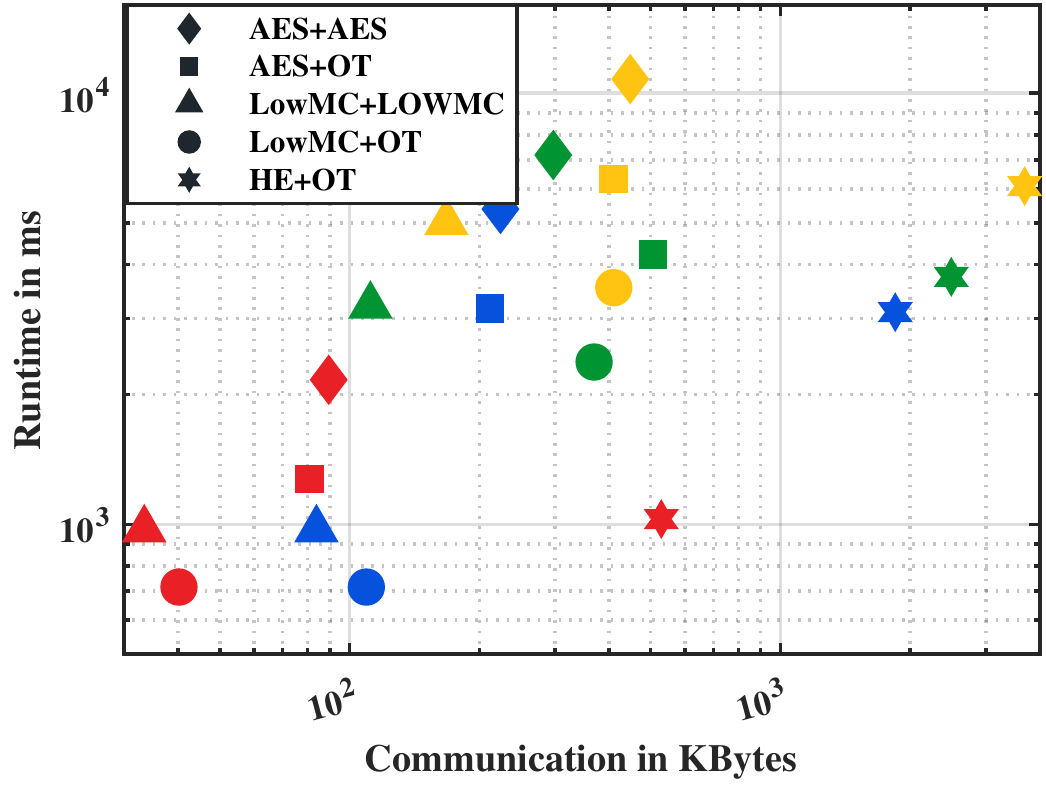}}
\subfigure[Trade-off under WAN (40Mbps/80ms)] {\label{fig:tradeoff:c}
\includegraphics[height=1.23in,width=1.63in]{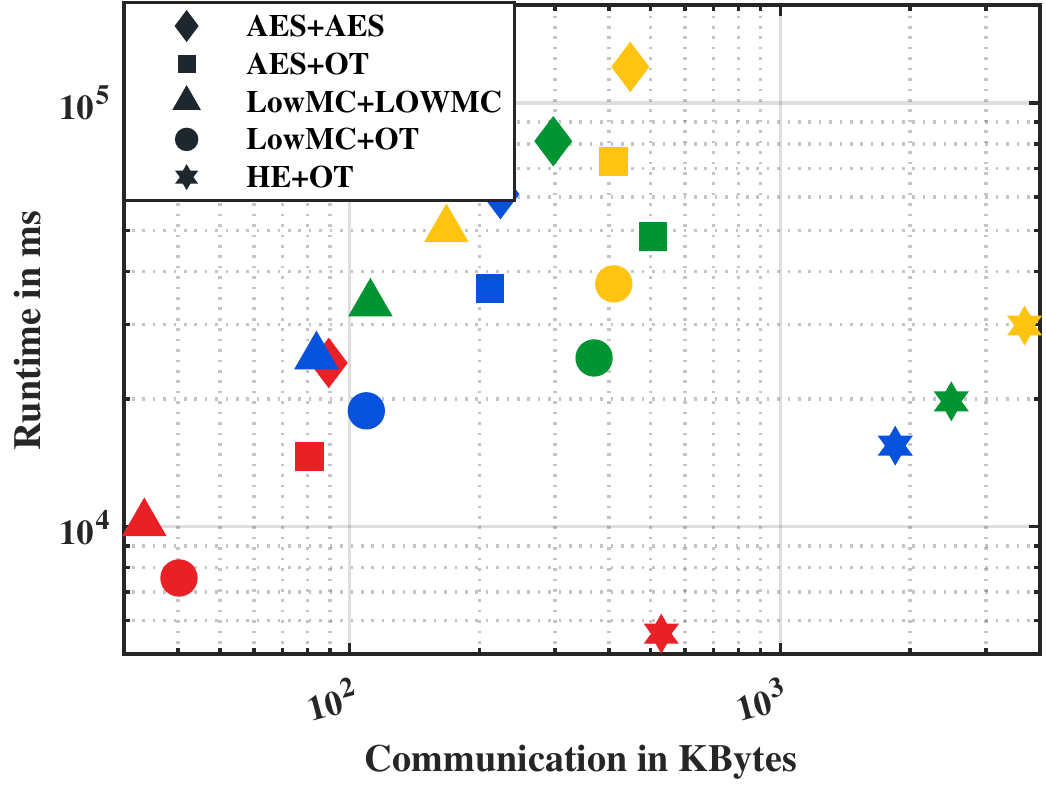}}
\caption{Trade-offs between Online Communication (x-axis) and Online Runtime (y-axis). Each figure shows the online complexities including communication and runtime. Note that diamond, square, triangle, round and six-pointed star represent AES$+$AES, AES$+$OT, LowMC$+$LowMC, LowMC$+$OT and HE+OT protocol, respectively. The shape filled by red, blue, yellow and green are Linnerud, digits, Boston and MNIST, respectively. Both $x$-axis and $y$-axis are in logarithm scale.
}
\label{fig:tradeoff}
\end{figure*}

\noindent\textbf{Trade-off.} 
We report the concrete trade-off between our PRF-based protocol and HE-based protocol. Since OT/PRF/HE can be used to instantiate SOS protocol, we investigate the differences when using different SOS protocols and give the corresponding experiments. To thoroughly examine the PRF-based protocol, we also implement AES as the underlying PRF. Specifically, we use A+B to denote the tree node is selected by A-based SOS protocol, and the attribute is obliviously shared by B-based SOS protocol.

We remove wine, breast, spambase and diabetes, which share similar parameters as the selected trees. 
Fig.~\ref{fig:tradeoff} shows the trade-off in three different network settings. 
In the scenarios with low network latency~(\eg, IoT), LowMC+OT can efficiently handle small trees like Linnerud. 
In the cases of deep trees with thousands of nodes, LowMC+LowMC shows less communication cost while LowMC+OT saves roughly 50\% runtime. Thus, in the situation where the computation performance matters more, LowMC+OT can be adopted to provide reasonable online computation overhead and communication cost. The reported runtime/communication of HE+OT under the LAN setting is significantly high than other PRF-based protocols.
With latency increasing, AES-based protocols suffer more than LowMC involved protocols, while HE+OT shows its advantages in runtime, as seen from Fig.~\ref{fig:tradeoff:b}. 
Under WAN with 80ms high network latency, the online running time for small trees by LowMC+LowMC is around 50\% higher than LowMC combining OT. 
This gap becomes progressively smaller as the tree depth increases. 
Under this setting, HE+OT is the first choice when considering time-consuming.

\begin{figure}[!tp]
\centering
    \subfigure[Total Communication] {\label{fig:scalability:a} \includegraphics[height=1.2in,width=1.6in]{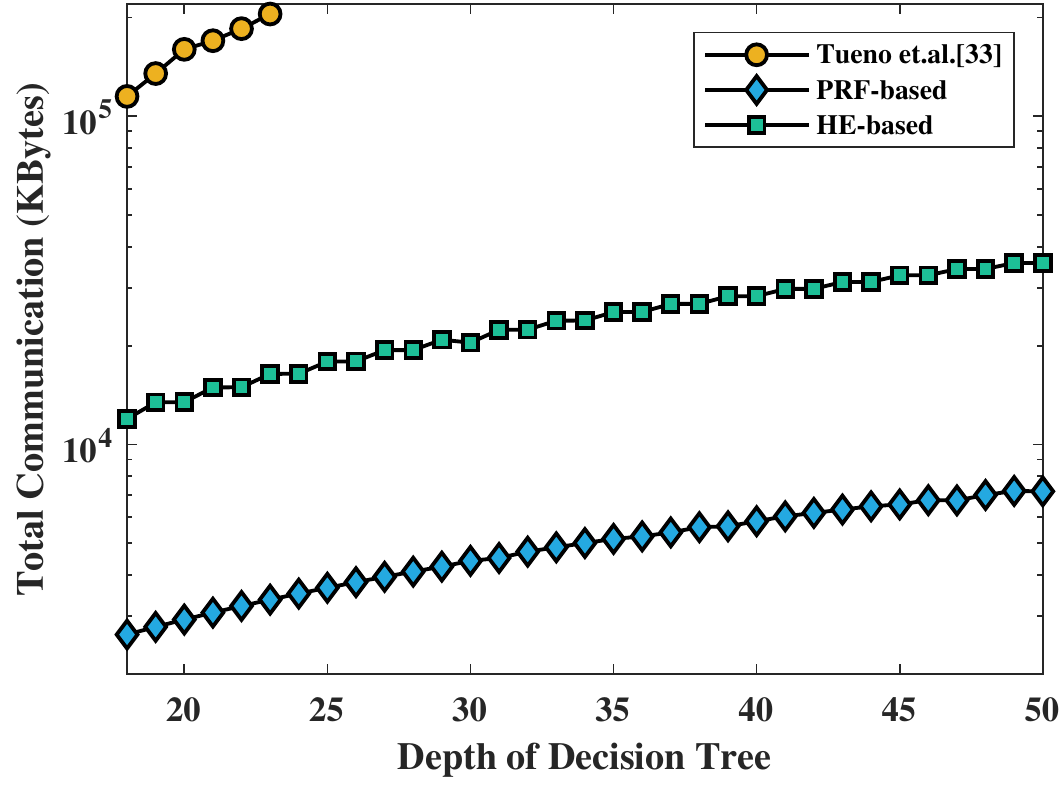}}
    \subfigure[Total Running Time]{\label{fig:scalability:b} \includegraphics[height=1.2in,width=1.6in]{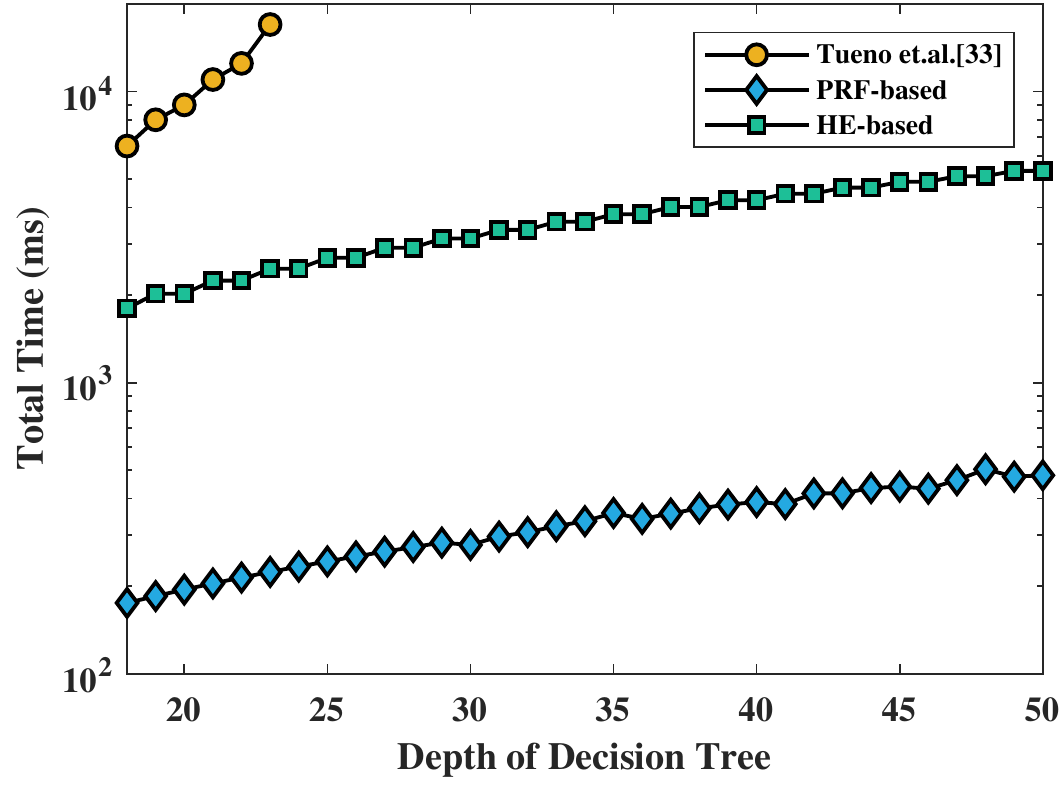}}
\caption{Total Communication and Runtime Cost (LAN setting). Note that $y$-axis is in logarithm scale. }
\label{fig:scalability}
\end{figure}

\noindent\textbf{Scalability for High-depth Trees}. 
We report the scalability of our PDTE protocols and compare them with the ORAM-based PDTE protocol~\cite{tueno2019private} in LAN setting.
Unfortunately, Tueno~\etal~\cite{tueno2019private} only report results for trees with depth up to 23 since their experiments ran out of memory.
The experiments of our protocols are done for trees with the depth ranging from 18 to 50.
For tree's size, we follow the same setting of \cite{tueno2019private} by letting $m = 25d$. 
We perform the evaluation in the LAN setting as shown in Fig.~\ref{fig:scalability}. 
As we can see, our HE-based protocol requires $9 \times$ to $16 \times$ less communication and $4 \times$ to $7 \times$ less running time compared with \cite{tueno2019private}.
Our PRF-based protocol requires $43 \times$ to $60 \times$ less communication and $26 \times$ to $54 \times$ less running time compared with \cite{tueno2019private}.
Therefore, our PDTE protocols are more scalable for evaluation over large trees. 
\vspace{-1mm}
\section{Related Work} \label{section:related}
There are many two-party PDTE protocols in the literature~\cite{brickell2007privacy,barni2009secure,bost2015machine,kiss2019sok,wu2016privately,tai2017privacy}. 
Among them, a few works~\cite{tueno2019private,joye2018private,ma2021let} achieve sublinear cost. In the following, we summarize these works separately. 

\noindent\textbf{Linear-cost Protocols.}
Brickell \etal~\cite{brickell2007privacy} pack each tree node into a GC circuit and then transmit the encrypted tree itself to the feature provider. With the help of HE and OT, the feature provider can perform oblivious evaluation by herself. However, the security requires that the encrypted tree should be refreshed for each tree evaluation, incurring linear costs. 
This work was later optimized by Barni \etal~\cite{barni2009secure}. The communication cost is saved by only sending encrypted internal nodes rather than the whole tree. However, it is still linear to the tree size. 
Bost \etal~\cite{bost2015machine} express the decision tree as a high-degree polynomial and encrypt it using an expensive Leveled Fully Homomorphic Encryption~(Leveled-FHE) scheme. 
Wu \etal~\cite{wu2016privately} propose a cheaper protocol by only relying on OT and AHE. Yet, the construction requires the tree holder to pad the tree to be complete in order to hide the tree structure. This padding strategy incurs massive communication and computation overhead, especially for deep-but-sparse trees. 
Subsequent PDTE protocols \cite{kiss2019sok, de2017efficient,liu2019secure} follow the same approach. To avoid padding, Tai \etal~\cite{tai2017privacy} propose a novel path cost mechanism using AHE. This protocol performs better for sparse trees but still runs at linear cost. 
Kiss \etal~\cite{kiss2019sok} systematically compare existing PDTE protocols. They mix HE and GC in different evaluation phases and report their concrete efficiency. Same as previous works, they trade efficiency for privacy by doing comparisons for all decision nodes, resulting in linear computation/communication. 
Ideally, the best solution is to perform only the necessary comparisons meanwhile hiding the decision path.

\noindent\textbf{Sublinear-cost Protocols.}
Three up-to-date work~\cite{joye2018private,tueno2019private,ma2021let} consider sub-linear decision tree protocols.
Joye and Salehi~\cite{joye2018private} reduce the number of secure comparisons to $d$. The comparison is based on DGK protocol~\cite{damgaard2007efficient} using AHE. In tree level $l$, they employ 1-out-of-$2^l$ OT to obliviously select an AHE encrypted tree node. In the end, the involved OT incurs $O( 2^d-1)$ communication in total. Thus, this PDTE protocol is only sublinear in computation.
Tueno \etal~\cite{tueno2019private} organize a decision tree as an array and build an oblivious array indexing (OAI). Such an interactive OAI allows the participants to pick the desired tree node and its corresponding attribute obliviously. 
OAI can be instantiated utilizing GC, OT or Oblivious RAM (ORAM). The first two OAIs can only realize sublinear complexity on the feature provider side since they also require OT to transfer among $2^d$ nodes, same as~\cite{joye2018private}. 
If employing ORAM, it takes $O(d^4)$ communication cost and requires $d^2$ (\eg, complete tree) rounds.

In order to further reduce communication cost, in protocol~\cite{ma2021let}, the tree holder encrypts the tree and sends it to the feature provider. In each tree level, there is an OT and a comparison between both parties. The feature provider searches local encrypted tree for the next node after comparison. Since the authors move the most expensive oblivious selection operations to feature provider's local computation, their protocol is very efficient in terms of computation. Nevertheless, we notice that this searching property means this tree cannot be reused across evaluations because the feature provider can learn some information from memory access patterns during different evaluations.
Accordingly, to reach the genuine PDTE when using~\cite{ma2021let}, toward every evaluation, the decision tree model should be re-randomized and transmitted to the feature provider, causing linear complexity in terms of communication and computation.

\noindent\textbf{Outsourced Protocols.} 
Some also try to gain more efficiency with the help of cloud servers, which is called outsourced PDTE protocols~\cite{liang2019efficient, ma2021let,ji2021uc,zheng2019towards}. They aim to use outsourced cloud servers to release the heavy burden from the tree holder and the feature provider. However, most of them suffer linear complexities~\cite{liang2019efficient, zheng2019towards} or leak more information~\cite{liang2019efficient} than prior two-party protocols. 
\section{Conclusion} \label{section:conclusion}
In this paper, we study how to design sublinear-communication PDTE protocols with improved efficiency.
We first propose two communication-efficient shared oblivious selection (SOS) protocols with different trade-offs.
By combining these SOS protocols with secure computation and a tree encoding strategy, we propose two PDTE protocols both with sublinear communication efficiency.
Our experiments show our protocols are efficient and practical. 
As future research, we will extend our techniques to other privacy-preserving machine learning protocols. 
\section*{Acknowledgment}
We thank the anonymous reviewers for insightful comments and suggestions. 
Bai and Russello would like to acknowledge the MBIE-funded programme STRATUS (UOWX1503) for its support and inspiration for this research.
This research is supported by the National Research Foundation, Singapore under its Strategic Capability Research Centres Funding Initiative. Any opinions, findings and conclusions or recommendations expressed in this material are those of the author(s) and do not reflect the views of National Research Foundation, Singapore.

\bibliographystyle{ACM-Reference-Format}
\bibliography{ref}

\appendix
\section{Security Proofs}

\subsection{Proof of Theorem~\ref{theorem:1}}\label{appendix::1::a}
\textbf{Security against corrupted $\sender$}. We construct a simulator $\SIM_{\rm s}$ as follows. During setup, $\SIM_{\rm s}$ runs $\sender$'s setup protocol except with the following exception: $\SIM_{\rm s}$ randomly samples $\CT[i] \xleftarrow{\$} \{0,1\}^{\Vbit}$ for all $i \in [0, m)$.
As for selection protocol, $\SIM_{\rm s}$ is provided with input $\boolshare{\idx}_{\rm s}$ and output $e$ (\ie, a PRF output share). It runs $\sender$'s selection protocol except with the following exceptions: $\SIM_{\rm s}$ runs simulator for $\Fpre$ to simulate the view of weight-1 bit vector generation.
Also, $\SIM_{\rm s}$ runs simulator for $\Fpre$ to simulate the view of shared PRF protocol. 
Finally, $\SIM_{\rm s}$ outputs $\sender$'s view.
We prove the simulated view is indistinguishable from real-world execution via a sequence of hybrid games. 

\begin{itemize}
    \item $\HYBRID_{\rm 0}$: Outputs $\sender$'s view in the real-world protocol. 

    \item $\HYBRID_{\rm 1}$: Same as $\HYBRID_{\rm 0}$ except $\SIM_{\rm s}$ randomly samples $\CT[i] \xleftarrow{\$} \{0,1\}^{\Vbit}$ for all $i \in [0, m)$.
    By the security of $\prf$, $\HYBRID_{\rm 1}$ is computationally indistinguishable from $\HYBRID_{\rm 0}$. %Otherwise, we can construct an adversary to distinguish $\prf$ from a random function with non-negligible probability. 

    \item $\HYBRID_2$: Same as $\HYBRID_{\rm 1}$ except that $\SIM_{\rm s}$ runs simulator for $\Fpre$ to generate $\sender$'s view in the shared weight-1 bit vector protocol.
    In particular, the simulator for $\Fpre$ is provided with input $n$ and output $\boolshare{\BV}_{\rm s}, \boolshare{\idx}_{\rm s}$, and in the end it generates a simulated view.  
    By the security of shared weight-1 bit vector protocol, $\HYBRID_2$ and $\HYBRID_{\rm 1}$ are indistinguishable.

    \item $\HYBRID_3$: Same as $\HYBRID_2$ except that $\SIM_{\rm s}$ runs simulator to generate $\sender$'s view in shared PRF protocol.
    Note that for all other view from running secret-shared computation, $\SIM_{\rm s}$ can simulate it by randomly sampling shares, which is identically distributed in both worlds.
    However, $\SIM_{\rm s}$ must ensure all prior simulated view be consistent with the final output, which is $\boolshare{\M[\idx][j]}_{\rm s}$.
    As such, $\SIM_{\rm s}$ computes $f_j^{*} \leftarrow \boolshare{\M[\idx][j]} \oplus  \boolshare{\CT[\idx][j]}^{*}$ for $j \in [0, \NumBlocks)$, it then invokes the simulator for $\Fpre$ over $(\boolshare{\{\idx||j}^{*}_{\rm s}\}_{j \in [0, \NumBlocks-1)}, \{f_j^{*}\}_{j \in [0, \NumBlocks)})$ to simulate $\sender$'s view in shared PRF protocol. 
    Since we are working in a hybrid model, such simulator must exist\footnote{A simulator who can simulate secret-shared XOR and AND computation will suffice to simulate any secure computation task.}. 
    $\HYBRID_3$ and $\HYBRID_2$ are indistinguishable by the security of shared PRF protocol.
\end{itemize}

%\smallskip 
\noindent\textbf{Security against corrupted $\receiver$}. 
Operations of $\sender$ and $\receiver$ are almost symmetric during the protocol, so our strategy for proving a corrupted $\receiver$ is almost same as the one we construct against corrupted $\sender$.
The proof for corrupted $\receiver$ is thus omitted.

\subsection{Proof of Theorem~\ref{theorem:sos-he}}\label{appendix::2::a}
    \textbf{Security against corrupted $\sender$}. 
    We construct a simulator $\SIM$ for corrupted $\sender$ as follows. 
    During setup, $\SIM$ runs $\sender$'s setup protocol except with the following exception: 
    $\SIM$ randomly samples $\CT[i] \xleftarrow{\$} \ZZ_{N^2}$ for all $i \in [0, m)$. 
    Given semantic security of Paillier encryption, the simulated ciphertexts are indistinguishable from the ones in real-world protocol.
    
    In the selection protocol, for the view from generating weight-1 bit vector, $\SIM$ calls the simulator for $\Fun_{\rm pre}$ to simulate corresponding view over randomly sampled $\boolshare{\BV}_{\rm s}$ and $\boolshare{\rdx}_{\rm s}$\footnote{Note that a secret share can be regards as a random number over its sharing domain. Then, the sampled random numbers as the simulated shares are indistinguishable from the real-world protocol.}; the simulation is perfect in hybrid mode. 
    The view of B2A conversion can also be simulated using existing simulator for secure computation, such a simulator always exists given the security of B2A protocol. 
    Next, $\SIM$ picks a random $\delta \rnd \ZZ_m$, the simulation is also indistinguishable: in the real-world protocol, $\idx$ and $\rdx$ are all random, then $\delta = \rdx -\idx~(\lmod~m)$ is also a random number in $\ZZ_m$.
    The simulator $\SIM$ performs simulation for the next B2A conversion protocol, similarly as it previously does for $\idx$ and $\rdx$. The simulated view of B2A is indistinguishable from real-world protocol.
    For the view of share conversion from additive sharing to multiplicative sharing, $\SIM$ first invokes the simulator for BMT generation in $\Fun_{\rm sprf}$ over a randomly sampled $a \rnd \ZZ_{N^2}$ and the share $\addshare{c} \rnd \ZZ_{N^2}$. 
    Next, $\SIM$ samples $\gamma, e, f$ and $\addshare{x\cdot \gamma^{-1}}$ randomly from $\ZZ_{N^2}$. 
    Note that in the real-world protocol, $e \leftarrow \gamma^{-1} - a~(\lmod~N^2)$ and $f \leftarrow \addshare{x}_{\rm r} - b~(\lmod~N^2)$.
    Given that $a, b, \gamma$ and $\addshare{x}_{\rm r}$ are all random, then $e$ and $f$ are random elements in $\ZZ_{N^2}$ as well. 
    However, there is a difference since the simulator samples $\gamma$ from $\ZZ_{N^2}$ rather from $\ZZ_{N^2}^*$, the simulated $\gamma$ can be accidentally sampled from $\ZZ_{N^2} \backslash \ZZ_{N^2}^*$. 
    However, this bad probability can only happens with probability of $\frac{1}{p}+\frac{1}{q}$, which is negligible in $\kappa$.  
    Next, $\SIM$ randomly samples $x_{\beta} \rnd \ZZ_{N^2}$ rather than $x_{\beta} \rnd \ZZ_{N^2}^*$, the probability to distinguish is negligible as we argued before.
    For $\addshare{m}_{\rm s}$, $\SIM$ will randomly sample it from $[0, 2^{{\Vbit}+\lambda+1}]$, this is statically closed to the real-world view for statistical parameter $\lambda$. 
    
    %\smallskip 
    \noindent\textbf{Security against corrupted $\receiver$}.  
    Simulator for a corrupted $\receiver$ can be constructed using the similar strategy as we constructed for $\sender$ since all operations are symmetric between $\sender$ and $\receiver$.
    Therefore, we omit the simulation for corrupted $\receiver$ in our proof.

%\smallskip 
\subsection{Proof of Theorem~\ref{theorem:2}}\label{appendix::3::a}
 
\textbf{Simulator for corrupted $\party_0$}.
The simulator $\SIM$ invokes the simulator of SOS setup protocol over $\ArrayTree$. 
Similarly, $\SIM$ invokes the simulator of SOS setup protocol over $\feature$. 
The simulation is perfect in the hybrid model. 
$\SIM$ randomly samples $\boolshare{t}_0, \boolshare{l}_0 \boolshare{r}_0, \boolshare{v}_0$ and $\boolshare{c}_0$ from $\ZZ_{2^\ell}$. It is straightforward to see that the simulated view is indistinguishable from the real-world execution.   

As for evaluation protocol, the simulator in the beginning does not need to simulate $\boolshare{\idx}$ since it shares 0, which is locally done by the parties.
Then $\SIM$ randomly samples $\boolshare{\mathsf{rst}}_0$. The above simulation is indistinguishable from the real-world view. 
Then $\SIM$ calls the simulator of SOS selection protocol over $\feature$.
In particular, $\SIM$ samples $\boolshare{v}_0$ and $\boolshare{\feature[v]}$ randomly, and invokes $\Fsos^{(\feature)}$ where $\boolshare{v}_0$ and $\boolshare{\feature[v]}$ are the input and the output, respectively.
Afterwards, $\SIM$ simulates the view of secure comparison, the view can be simulated as the secure computation is well-studied in existing work~\cite{demmler2015aby}.
Then $\SIM$ randomly samples $\boolshare{\idx} \in \ZZ_{2^{\ell}}$, the simulation is perfect since the share is randomly computed from secure computation.
$\SIM$ invokes the simulator for SOS selection protocol over $\ArrayTree$ with $\boolshare{\idx}_{\rm s}$ as the input and a random number $r \rnd (\ZZ_{2^{\Vbit}})^5$ as the output. 
In particular, $r$ is used to simulate $\sender$'s share of $\ArrayTree[\idx]$. 
The simulated view is indistinguishable from real-world view due to the security of SOS protocol. 
For all other operations that can be done locally, $\SIM$ can simulate trivially (since no view involves in these operations).

%\smallskip
\noindent\textbf{Simulator for corrupted $\party_1$}.
The idea of simulating view for a corrupted $\party_1$ is almost the same as the simulation for $\party_0$ because the PDTE protocol is essentially symmetric for the two parties. 
Therefore, we omit the proof for the corrupted  $\party_1$. 
% \end{proof}

%\smallskip 
\section{Correlated randomness generation}\label{appendix::2::c}
\subsection{BMT Generation} 
For a normal BMT  $(\addshare{a}, \addshare{b}, \addshare{c})$, $a$, $b$ and $c$ are all secret-shared among the parties. 
In our setting, we want to generate BMT $(a, b, \addshare{c})$ where $\party_0$ holds $(a, \addshare{c}_0)$ and $\party_1$ holds $(b, \addshare{c}_1)$.
In the following, we summarize two ways for generating such special BMTs using either AHE or OT.

%\smallskip 
\noindent\textbf{AHE-based approach \cite{demmler2015aby}}. 
BMTs can be generated from AHE, \eg, Paillier encryption.
The parties can explore the additive homomorphic property to share the multiplication result over $\ZZ_n$.
In Fig.~\ref{protocol:BMT-AHE}, we give a protocol for generating BMTs from Paillier encryption.

\begin{figure}%[!hp]
	\framebox{\begin{minipage}{0.95\linewidth}
		{\bf Parameters}: 
			BMT module $n$, computational security parameter $\kappa$; statistical security parameter $\lambda$; Paillier plaintext module $N= p\cdot q$, where $p$ and $q$ are primes.
			
		\begin{enumerate}
		    \item $\party_0$ generates a pair of Paillier public/private key pair $(\pk, \sk) \leftarrow \textsf{Gen}(1^{\kappa})$ and sends $\pk$ to $\party_1$.
		    
		    \item $\party_0$ chooses $a \rnd \ZZ_{n}$ and sends $x \leftarrow \textsf{Enc}_{\pk}(a)$ to $\party_1$. 
		    
		    \item $\party_1$ chooses $b \rnd \ZZ_{n}$ and $\rho \rnd [0, 2^{\lambda})$, sends $x' \leftarrow x^b \cdot \textsf{Enc}_{\pk}(r + \rho\cdot n)~(\lmod~N^2) $ to $\party_0$.
		    
		    \item $\party_0$ decrypts to get $\addshare{c}_0 \leftarrow \textsf{Dec}_{\sk}(x')~(\lmod~n)$. 
		    
		    \item $\party_1$ sets $\addshare{c}_1 \leftarrow -r~(\lmod~n)$. 
		    
		\end{enumerate}
	\end{minipage}}
	\caption{BMT from AHE~\cite{demmler2015aby}}
	\label{protocol:BMT-AHE}
\end{figure}

\begin{figure}%[!hp]
	\framebox{\begin{minipage}{0.95\linewidth}
		{\bf Parameters}: 
			BMT module $n$, $n$'s bit size $\ell$; computational security parameter $\kappa$.
		
		%For $0 \le i < \ell$:
		\begin{enumerate}
		    \item $\party_0$ chooses $a \rnd \ZZ_{n}$ and decomposes $a$ to its boolean form $(a_{\ell-1}, \cdots, a_1, a_0)$ such that $a = \sum_{i=0}^{\ell-1}2^i\cdot a_i$.
		    
		    \item For $0 \le i < \ell$: 
		    \begin{enumerate}
		        \item $\party_1$ chooses $r_i \rnd \ZZ_n$ and computes two messages $(m_i^0 = r_i, m_i^1 = r_i + 2^i\cdot b~(\lmod~n))$;
		    
		        \item $\party_0$ and $\party_1$ invoke 1-out-of-2 OT functionality $\Fun_{\rm OT}$ where $\party_1$ sends $(m_i^0, m_i^1)$ and $\party_0$ sends $a_i$. In the end, $\party_0$ receives $m_i^{a_i}$.
		    \end{enumerate}
		    \item $\party_0$ sets $\addshare{c}_0 \leftarrow \sum_{i=0}^{\ell-1} m_i^{a_i}~(\lmod~n)$, and $\party_1$ sets  $\addshare{c}_1 \leftarrow \sum_{i=0}^{\ell-1}-r_i~(\lmod~n)$. 
		\end{enumerate}
	\end{minipage}}
	\caption{BMT from OT~\cite{demmler2015aby}}
	\label{protocol:BMT-OT}
\end{figure}

%\smallskip 
\noindent\textbf{OT-based approach \cite{demmler2015aby}}
A BMT $(a, b, \addshare{c})$ can be generated by using OTs as shown in Fig.~\ref{protocol:BMT-OT}.

\subsection{WBV Generation from FSS}
Boyle~\etal~\cite{gilboa2014distributed,boyle2015function,boyle2016function} formalize a new cryptographic primitive called \emph{Function Secret Sharing}~(FSS), and gave concrete constructions for useful functions.
We begin with formally defining \emph{Function Secret Sharing} for two parties.

\begin{definition}[Function Secret Sharing]
\label{Def:FSS}
	A two-party FSS scheme ${\rm \Pi_{\rm fss}} = (\textsf{\rm Gen}, \textsf{\rm Eval})$ consists of a pair PPT algorithms as follows:
	\begin{itemize}
		\item $\textsf{\rm Gen} (1^{\kappa}, f)$ is a key generation algorithm, which takes as input a security parameter $1^{\kappa}$ and a function description $f$, outputs a tuple of keys $(k_0^{\rm fss}, k_1^{\rm fss})$, each for one party.
		\item $\textsf{\rm Eval} (k_i^{\rm fss}, x)$ is an evaluation algorithm, which on input a key $k_i^{\rm fss}$ for party $P_i$~($i \in \{0, 1\}$), and an evaluation point $x \in \{0, 1\}^\ell$, outputs a group element $y_i \in \mathbb{G}$ as the share of $f(x)$ for $P_i$.
	\end{itemize}
\end{definition}

\begin{definition}[Security of FSS] 
\label{Def:FSSsecurity}
	A secure two-party FSS satisfies the following requirements:
	\begin{itemize}
		\item $\textsf{\rm Correctness}$: for all function $f: \{0, 1\}^{\ell} \rightarrow \mathbb{G}$ and every $x \in \{0, 1\}^n$, if $(k_0^{\rm fss}, k_1^{\rm fss}) \leftarrow \textsf{\rm Gen} (1^{\kappa}, f)$ then ${\rm Pr} [\textsf{\rm Eval}(k_0^{\rm fss}, x)+\textsf{\rm Eval}(k_1^{\rm fss}, x)=f(x)] = 1$.
		
		\item $\textsf{\rm Secrecy}$: For every corrupted $P_i$ and every sequence of function $f_1, f_2,...$, there exists a PPT simulator $\SIM$ such that 	for $i \in \{0, 1\}$:%\note{$f_1, f_2,...f_k$}
		\[\{ k_i^{\rm fss}:(k_0^{\rm fss}, k_1^{\rm fss}) \leftarrow \textsf{\rm Gen}(1^{\kappa}, f_\kappa) \}_{\kappa \in \mathbb{N}} \stackrel{c}{\equiv} \{ \SIM_i(1^{\kappa}, i, \mathbb{G}) \}_{\kappa \in \mathbb{N}}\]
	\end{itemize}
\end{definition}

\begin{definition}[Point Function]\label{def:pointfun}
	A point function is a function $f_{\alpha, \beta}(x): \{0,1\}^{\ell} \rightarrow \mathbb{G}$ where $\mathbb{G}$ is an abelian group such that \begin{equation}
	f_{\alpha, \beta}(x) = \left\{
	\begin{array}{ll}
	\beta, & {\rm if} \; x = \alpha \\
	0, & {\rm otherwize} \\
	\end{array}
	\right.
	\end{equation}
\end{definition}

Our WBV is based on a point function defined in Definition~\ref{def:pointfun}, which can be shared using FSS with the key size of $O(\kappa \log m)$.
A trusted dealer can generate the keys for a point function with $O(\log m)$ PRG evaluations. 
In \cite{doerner2017scaling}, they use two-party computation to generate FSS keys, removing the trusted-dealer assumption. 
Indeed, the technique in \cite{doerner2017scaling} can be directly used for pre-processing WBV.
Specifically, each party $\party_i$ just locally sample a random share $\boolshare{\rdx}_i \in \ZZ_2^{\ell}$, and then run the two-party FSS key generation protocol of \cite{doerner2017scaling} to compute the keys. 
Each party will hold a FSS key after secure computation, and then each party can evaluate his key over $i \in [0,m)$ to generate his own share of a WBV. 
By definition of point function, the parties will only share $1$ at $\rdx$, and 0 for any $i \ne \rdx$. 

\end{document}